\documentclass{article}
\usepackage[preprint]{neurips_2026}
\setcitestyle{authoryear,round,semicolon,aysep={,},yysep={,}}
\usepackage[T1]{fontenc}
\usepackage[utf8]{inputenc}
\usepackage{microtype}

\usepackage{amsmath,mathtools,amssymb,amsthm}
\usepackage{algorithm}
\usepackage{algpseudocode}

\usepackage{graphicx}
\usepackage{wrapfig,needspace}
\graphicspath{{figures/}}
\usepackage{booktabs,multirow,array}
\usepackage{capt-of}
\usepackage{rotating}
\usepackage{xcolor}
\usepackage[normalem]{ulem} 

\usepackage[hyperfootnotes=false,colorlinks=true,linkcolor=blue!55!black,citecolor=blue!55!black,urlcolor=blue!55!black]{hyperref}
\usepackage{url}
\hypersetup{pdftitle={DynaCrys: Crystal Generation with Dynamic Space-Group Diffusion},
pdfauthor={Zhuotao Jin, Xiaoyun Wang, Nicholas Brawand, Roman Zubatyuk, Atul Thakur, Eric Qu, Boris Kozinsky, Justin Smith}}

\newcommand{\ours}{DynaCrys}
\newcommand{\ourssmall}{DynaCrys-small}
\newcommand{\maintablefont}{\fontsize{8pt}{10pt}\selectfont}

\theoremstyle{definition}
\newtheorem{proposition}{Proposition}

\title{\ours{}: Crystal Generation with Dynamic Space-Group Diffusion}
\author{Zhuotao Jin$^{\,1,2,3,4,\dagger}$, \ Xiaoyun Wang$^{\,1}$, \ Nicholas Brawand$^{\,5}$, \ Roman Zubatyuk$^{\,1}$,\\
\textbf{Atul Thakur$^{\,1}$, \ Eric Qu$^{\,1,6}$, \ Boris Kozinsky$^{\,4,7}$, \ Justin Smith$^{\,1,\dagger}$}\\[6pt]
{\small $^{1}$NVIDIA}\\
{\small $^{2}$Center for Computational Science and Engineering, Massachusetts Institute of Technology}\\
{\small $^{3}$Department of Materials Science and Engineering, Massachusetts Institute of Technology}\\
{\small $^{4}$John A. Paulson School of Engineering and Applied Sciences, Harvard University}\\
{\small $^{5}$The MITRE Corporation}\\
{\small $^{6}$University of California, Berkeley}\\
{\small $^{7}$Robert Bosch Research and Technology Center}
}
\date{}

\begin{document}
\typeout{MANUSCRIPT-MODE: preprint}
\maketitle
{\let\thefootnote\relax\footnotetext{$^{\dagger}$Corresponding authors.}}

\begin{abstract}
The search for new crystalline materials spans an enormous compositional and structural
space. Generating candidates in this space requires modeling the relationships between
elemental composition, crystallographic symmetry, and crystal geometry. The space group
specifies the crystal's symmetry and determines the admissible Wyckoff positions and
geometric degrees of freedom; fixing it in advance prevents this global symmetry choice
from being revised using information from evolving Wyckoff occupations and elemental
composition. We introduce \ours{}, the first framework for complete
crystal generation in which the space group co-evolves with Wyckoff occupations and
elements. We develop physics-informed space-group diffusion that explicitly
incorporates crystallographic group--subgroup relations, and a legality-constrained stochastic
decoder that jointly samples compatible space groups, Wyckoff occupations, and elements,
followed by symmetry-constrained geometry generation. In large-scale evaluations
on standard benchmarks, \ours{} achieves
state-of-the-art performance in symmetry-aware crystal generation while supporting
fast sampling.
\end{abstract}

\section{Introduction}
\label{sec:intro}

Discovering new crystalline materials requires searching a vast space of elemental
compositions and atomic arrangements. Generative models offer a way to propose
candidates directly, but the variables that
define a crystal are tightly coupled. Its space group $G$ determines the admissible
Wyckoff positions and their geometric degrees of freedom.
The space group, Wyckoff occupations, and their elements define the
crystal's protostructure---its symbolic framework. Lattice parameters and atomic
coordinates specify its geometry within the corresponding symmetry constraints.

\begin{wraptable}[12]{R}{272pt}
\setlength{\abovecaptionskip}{0pt}
\centering
\caption{Crystallographic symmetry across models.}
\label{tab:method-comparison}
\begingroup%
\small%
\setlength{\tabcolsep}{2pt}%
\renewcommand{\arraystretch}{1.08}%
\begin{tabular}{@{}lccc@{}}
\toprule
Methods & \begin{tabular}{@{}c@{}}Symmetry-\\aware\end{tabular} & \begin{tabular}{@{}c@{}}Dynamic\\$G$\end{tabular} & \begin{tabular}{@{}c@{}}Group--subgroup\\structure\end{tabular} \\
\midrule
\begin{tabular}{@{}l@{}}MatterGen, FlowMM, DiffCSP\end{tabular} & $\times$ & $\times$ & $\times$ \\
\midrule[0.3pt]
\begin{tabular}{@{}l@{}}DiffCSP++, SymmCD, WyFormer\\SGEquiDiff, WyckoffDiff\end{tabular} & $\pmb{\checkmark}$ & $\times$ & $\times$ \\
\midrule
\ours{} (ours) & $\pmb{\checkmark}$ & $\pmb{\checkmark}$ & $\pmb{\checkmark}$ \\
\bottomrule
\end{tabular}%
\endgroup%

\end{wraptable}

In crystal generation, models such as MatterGen \citep{mattergen} and FlowMM
\citep{flowmm} directly model lattice parameters,
atomic coordinates, and elements, with space groups determined post hoc from the generated
structures. In contrast, symmetry-aware models explicitly use space groups,
Wyckoff occupations, or site symmetries to constrain generation
(Table~\ref{tab:method-comparison}). These models, however, select and fix the space
group before generating the remaining variables, restricting exploration
to configurations compatible with that group. Joint generation can instead use information
from the evolving occupations and elemental composition to update the space-group choice.

\begin{figure}[htb]
\centering
\includegraphics[width=\linewidth]{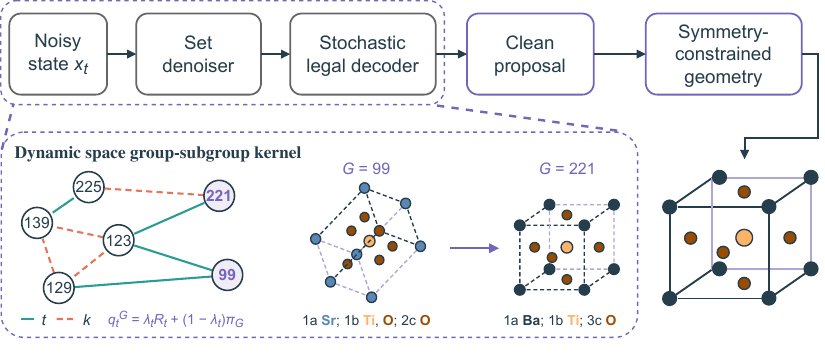}
\caption{Overview of \ours{}: joint generation of the space group, Wyckoff
occupations, and elements, followed by symmetry-constrained geometry generation.}
\label{fig:overview}
\end{figure}

We introduce \ours{}, a framework for complete crystal generation in which the space
group jointly evolves and is dynamically revised with Wyckoff occupations and elements through
diffusion. The physics-informed
space-group diffusion kernel incorporates crystallographic group--subgroup relations
\citep{barnighausen}, which describe changes in symmetry constraints
and their associated degrees of freedom. A legality-constrained stochastic decoder
then samples compatible symbolic states. Conditioned on the final
state, a symmetry-constrained geometry model generates lattice parameters and atomic
coordinates to produce a complete crystal (Figure~\ref{fig:overview}).

Our contributions are:
\begin{list}{\labelitemi}{%
\setlength{\leftmargin}{0pt}%
\setlength{\itemindent}{1em}%
\setlength{\labelwidth}{0.5em}%
\setlength{\labelsep}{0.5em}%
\setlength{\itemsep}{2pt}%
}
\item \emph{Joint crystal generation with a dynamic space group.} We introduce the first complete crystal
generator to jointly evolve the space group, Wyckoff occupations, and elements during generation.
\item \emph{Physics-informed space-group diffusion.} We incorporate the group-theoretic
structure of crystal symmetries into the space-group diffusion kernel, using
crystallographic group--subgroup relations as a structured prior for joint generation.
\item \emph{Legality-constrained stochastic decoding.} We develop a joint stochastic
decoder. It can sample according to model-predicted probabilities conditioned on space-group
and occupation constraints, without costly summation over all legal configurations.
\item \emph{Material discovery and validation.} We demonstrate state-of-the-art generation
performance among symmetry-aware crystal generative models on MP-20 and MPTS-52.
Independent evaluations using multiple machine-learning interatomic potentials and
large-scale first-principles calculations consistently confirm that \ours{} generates
a higher proportion of stable, unique, and novel crystals.
\end{list}

\section{Related Work}
\label{sec:related}

\paragraph{Crystal generation without explicit space-group modeling.}
CDVAE \citep{cdvae}, DiffCSP \citep{diffcsp}, and MatterGen \citep{mattergen}
use diffusion-based generative models, while FlowMM \citep{flowmm} uses Riemannian
flow matching. These methods directly model lattice parameters, atomic coordinates,
and elements. They use periodic representations and geometric invariance or
equivariance to handle global translations and rotations. Their space groups are
identified post hoc from the generated structures rather than represented
as explicit generative variables.

\paragraph{Crystal generation with explicit crystallographic symmetry.}
Explicitly using space groups, Wyckoff positions, or site symmetries allows generative
models to enforce crystallographic constraints directly.
Existing methods determine orbit occupations through different mechanisms.
Template-based DiffCSP++ \citep{diffcsppp} directly obtains space groups and occupation
patterns from training crystal data, then assigns elements to symmetry orbits and generates geometry.
Other methods learn to generate occupations: SymmCD \citep{symmcd} determines
Wyckoff positions by generating site symmetries; WyFormer \citep{kazeev2025wyckoff}
and SGEquiDiff \citep{sgequidiff} generate occupations and elements autoregressively;
and WyckoffDiff \citep{wyckoffdiff} diffuses discrete occupation variables conditioned
on the space group. These methods select and fix the space group before generating
the remaining variables. \ours{} jointly updates the space
group, Wyckoff occupations, and elements through diffusion guided by group--subgroup
relations, allowing the space-group choice to change during generation
(Table~\ref{tab:method-comparison}).

\paragraph{Constrained decoding for crystal generation.}
Existing crystal generative models handle constraints in different ways.
SGEquiDiff \citep{sgequidiff} prevents repeated occupation of 0D Wyckoff
positions during autoregressive sampling; WyckoffDiff \citep{wyckoffdiff} enforces
this restriction through its occupation representation; and SymmCD \citep{symmcd}
obtains valid site symmetries and coordinates through projection.
\ours{} uses legality-constrained joint stochastic decoding to sample compatible states
because it needs to coordinate the dynamic space group and other variables,
ensuring that the predicted orbit count matches the number of selected orbit occupations
and that occupation-capacity and atom-count constraints are satisfied.

\section{Preliminaries: Space Groups and Wyckoff Orbits}
\label{sec:crystal}
\label{sec:vocab}

A periodic crystal consists of a lattice basis $L\in\mathbb R^{3\times3}$ and atoms in one
unit cell, represented here by fractional coordinates $\mathbf f\in[0,1)^3$. Rather than
treating these atoms as independent points, a space group $G$ organizes them into symmetry
orbits. There are 230 three-dimensional space-group types \citep{ITA}. Each operation
$g=(R_g,\mathbf t_g)$ acts on fractional coordinates as
$g\cdot\mathbf f=(R_g\mathbf f+\mathbf t_g)\bmod 1$. The orbit of a representative
coordinate is the set of distinct symmetry-equivalent sites obtained by applying all
operations in $G$.

\begin{wrapfigure}[15]{r}{3.94in}
\centering
\includegraphics[width=\linewidth,trim=0bp 1.2bp 0bp 0.25bp,clip]{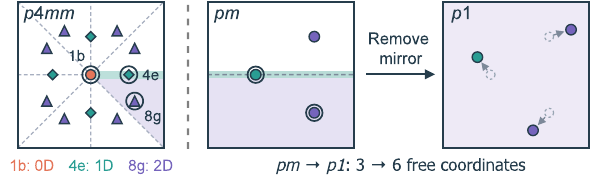}
\caption{Schematic of Wyckoff orbits and symmetry constraints.
Left: $p4mm$ orbits with circled representatives; matching markers denote equivalent
sites (inspired by \citealp{diffcsppp}). Right: orbit splitting under $pm\rightarrow p1$,
with dashed circles marking the original positions.}
\label{fig:background}
\end{wrapfigure}

A Wyckoff position describes a family of symmetry orbits. Each Wyckoff position
corresponds to a row in the space-group table, indexed by $r\in\mathcal R_G$.
The row specifies the orbit multiplicity $\mu_{G,r}$ (the number of equivalent sites
in an orbit) and the number of free coordinates $\operatorname{dof}_{G,r}$.
Choosing values for the free coordinates selects a particular orbit; a row with no
free coordinates already determines the orbit. For example, $r$ can denote the row
labeled $4e$ in Figure~\ref{fig:background} (left). The four matching diamonds form
one orbit of this Wyckoff position. Here, $4$ gives the multiplicity and $e$ identifies
the Wyckoff position within $G$ \citep{ITA}. Varying the free coordinate in this example
can produce another orbit represented by the same row.

A symbolic orbit occupation specifies a row $r$ and an element $Z$, leaving the
free-coordinate values unspecified. Each occupation therefore contributes $\mu_{G,r}$
atoms of element $Z$. Rows with free coordinates may be occupied more than once.
Together with $G$, these occupations define the
protostructure; geometry generation supplies the lattice parameters and remaining
free coordinates.

The vocabulary $\mathcal R_G$ is specific to the space group. Wyckoff labels are defined
within each space group. Changing $G$ changes the available rows, their multiplicities, and their
free-coordinate parameterizations. Group--subgroup relations organize these changes. A
maximal relation connects two space groups with no intermediate space group. Every such
step is either translationengleiche ($t$), which preserves the translation lattice while
reducing point group symmetry, or klassengleiche ($k$), which retains the point group
while reducing translational symmetry \citep{ITA}. Restricting $G$ to a subgroup $H$ can
split one $G$-orbit into several independently parameterized $H$-orbits.
In the illustrated two-dimensional $pm\rightarrow p1$ transition
(Figure~\ref{fig:background}, right), the three atoms can then vary independently,
increasing the number of free coordinates from $3$ to $6$.

Consequently, revising the space group during symbolic generation does more than update a
categorical label: it changes the admissible orbit representation and the continuous degrees of
freedom available to the subsequent geometry stage. Section~\ref{sec:method} therefore treats
$G$, the orbit occupations, and their elements as coupled components of one evolving symbolic state.

\section{Dynamic Space-Group Crystal Generation}
\label{sec:method}

\ours{} separates crystal generation into two stages. A symbolic diffusion model first
samples a protostructure containing the space group, Wyckoff occupations, and
their elements. These variables evolve jointly during this stage.
Conditioned on this symbolic state, a symmetry-constrained geometry model
then generates the remaining lattice parameters and free coordinates.

\subsection{Protostructure state and shared symmetry representation}
\label{sec:state}

Let $\mathcal G$ denote the space-group vocabulary. A
protostructure $P$ \citep{wren} is
\begin{equation}
P=\bigl(G,\;K,\;\{(Z_i,r_i)\}_{i=1}^{K}\bigr),
\label{eq:state}
\end{equation}
where $G\in\mathcal G$ is the space group, $K$ is the number
of orbit occupations, $r_i\in\mathcal R_G$ is a Wyckoff row, and $Z_i$ is
the element assigned to that occupation. The occupations form an unordered
multiset, the same row may occur more than once, and $P$ contains neither a lattice
nor free coordinates. The symbolic diffusion represents $P$ with a fixed-width padded clean state $x_0$.
At noise level $t$, the state is
\begin{equation}
x_t
=\bigl(G_t,\;K_t,\;\{(m_{i,t},Z_{i,t},y_{i,t})\}_{i=1}^{M}\bigr).
\label{eq:noisy-state}
\end{equation}
Here $m_{i,t}$ marks an active or padded slot, $Z_{i,t}$ is the corresponding
element or padding symbol, and $y_{i,t}\in\mathbb R^d$ is a noisy Wyckoff-row
vector. At $t=0$, the active slots encode the occupations of $P$ and the remaining
slots are padding. The noisy state $x_t$ need not decode directly to a legal
protostructure.

Because the row vocabulary $\mathcal R_G$ depends on $G$, \ours{} uses a shared symmetry
codebook containing space-group vectors $c_G$ and Wyckoff-row vectors $c_{G,r}$.
The codebook is pretrained from crystallographic descriptors. For an active clean slot,
$y_{i,0}=c_{G,r_i}$, linking its discrete row to the continuous row channel.
The row vectors also serve as the tied vectors of the
space-group-conditional row head. The geometry model is also conditioned on the space-group and row
codebook embeddings. At each diffusion step, $c_{G_t}$ represents
the current noisy space group $G_t$. The shared codebook therefore provides a common
representation of space-group-specific Wyckoff vocabularies across the symbolic and geometry
stages. Appendix~\ref{app:codebook} provides details of the codebook construction.

\subsection{Joint symbolic diffusion}
\label{sec:dyng}
\label{sec:denoiser}

\paragraph{Forward corruption.}
The symbolic forward process is a diffusion process defined by a family of marginal kernels
$q_t(x_t\mid x_0)$ \citep{d3pm}. Its space-group channel
mixes a structured transition with
a reset to the training distribution:
\begin{equation}
q_t^{G}(G_t\mid G_0)
=\lambda_t R_t(G_t\mid G_0)
+(1-\lambda_t)\pi_G(G_t).
\label{eq:gkernel}
\end{equation}
The transition $R_t$ diffuses over the undirected graph induced by maximal
translationengleiche and klassengleiche group--subgroup relations
on $\mathcal G$. This graph supplies the transition geometry of the
space-group corruption; the reset component makes the terminal distribution
independent of $G_0$.

After drawing $G_t$, the forward process samples $K_t$ and partitions the $M$
slots into survivors, births, and padding. Survivors retain a corrupted signal
from distinct clean occupations, whereas birth rows are drawn from the
vocabulary of the current $G_t$; the row vectors and elements are then
corrupted conditionally on this partition. This construction accommodates a
changing number of occupations and defines the same marginal kernel for
training-time corruption and sampling-time re-noising.
Further details on the channel distributions and noise schedules are provided in
Appendix~\ref{app:corruption}.

\paragraph{Set denoising network.}
The denoiser represents $x_t$ by one global token for $(G_t,K_t,t)$ and $M$
orbit-slot tokens containing $(m_{i,t},Z_{i,t},y_{i,t})$. A Transformer encoder
\citep{vaswani2017} processes these tokens without slot positional encodings,
making its slot outputs permutation equivariant. Global and slot-wise heads
predict the clean state as
\begin{equation}
p_\theta(G_0\mid x_t),\qquad
p_\theta(K_0\mid x_t),\qquad
\left\{
p_i^{\mathrm{act}}(\,\cdot\mid x_t),\;
p_i(r\mid G,x_t),\;
p_i(Z\mid G,r,x_t)
\right\}_{i=1}^{M}.
\label{eq:heads}
\end{equation}
Here $p_\theta(G_0\mid x_t)$ and $p_\theta(K_0\mid x_t)$ are the global
predictions of the clean space group and orbit count. For slot $i$,
$p_i^{\mathrm{act}}$ is the binary distribution over occupied ($1$) and padded ($0$)
states. The row head
$p_i(r\mid G,x_t)$ assigns probability only to valid rows $r\in\mathcal R_G$ and
uses their shared-codebook vectors as classifier weights. The element head
$p_i(Z\mid G,r,x_t)$ conditions its predictions on the selected $(G,r)$ through
feature-wise linear modulation (FiLM) \citep{film}
(Appendix~\ref{app:den-heads}).

\paragraph{Training objective.}
The symbolic model predicts $x_0$ with the set objective
\begingroup
\medmuskip=1mu
\thickmuskip=3mu
\begin{equation}
\mathcal L_{\mathrm{sym}}
=\mathcal L_G+w_K\mathcal L_K+w_m\mathcal L_{\mathrm{mask}}
+w_r\mathcal L_r+w_Z\mathcal L_Z
+s_{\mathrm{aux}}(
w_{\mathrm{cap}}\mathcal L_{\mathrm{cap}}
+w_N\mathcal L_N
+w_{\mathrm{comp}}\mathcal L_{\mathrm{comp}}
).
\label{eq:symbolic-loss}
\end{equation}
\endgroup
Here $\mathcal L_G$ and $\mathcal L_K$ are cross-entropies for the global
space-group and orbit-count predictions. A Hungarian assignment
\citep{carion2020detr} matches the $M$ predicted slots to the $K_0$ clean
occupations and $M-K_0$ padding targets before the slot losses are evaluated.
After matching, $\mathcal L_{\mathrm{mask}}$ is the active-state binary
cross-entropy over all slots, while $\mathcal L_r$ and $\mathcal L_Z$ are the row
and element negative log-likelihoods over occupied matches. Among the auxiliary
terms, $\mathcal L_{\mathrm{cap}}$ penalizes excess expected row occupancy, while
$\mathcal L_N$ and $\mathcal L_{\mathrm{comp}}$ match the expected
conventional-cell atom count and composition to the clean target. The sampler
separately enforces row-capacity limits and the fixed atom-count cap.
A separate fine-tuning stage refines the row-conditioned element head with regularization.
Appendix~\ref{app:denoiser}
gives the loss weights, exact assignment, auxiliary-loss definitions,
and optimization settings.

\subsection{Legality-constrained sampling}
\label{sec:decoding}

The heads in Eq.~\eqref{eq:heads} define a factorized clean-state distribution, but
the independently predicted orbit count and active states need not agree, and the
selected rows need not jointly satisfy the row-capacity and atom-count constraints.
Let $\mathcal X_{\mathrm{legal}}$ denote the corresponding legal set:
$G$ must lie in $\mathcal G$, exactly $K$ slots must be active, every selected row must
belong to $\mathcal R_G$, and the collection of occupations must satisfy the row-capacity
and atom-count constraints. The stochastic decoder targets the factorized distribution
conditioned on this set,
\begin{equation}
\widetilde p_\theta(x_0\mid x_t)
\propto
p_\theta^{\mathrm{fac}}(x_0\mid x_t)
\mathbf 1[x_0\in\mathcal X_{\mathrm{legal}}].
\label{eq:target}
\end{equation}
Here $p_\theta^{\mathrm{fac}}$ is the factorized distribution induced by the
heads in Eq.~\eqref{eq:heads}, $\mathbf 1[\cdot]$ is the legality indicator, and
$\widetilde p_\theta$ is the resulting distribution restricted to and
renormalized over $\mathcal X_{\mathrm{legal}}$.

The global constrained normalizer for Eq.~\eqref{eq:target} is the total probability
mass that the factorized model assigns to $\mathcal X_{\mathrm{legal}}$.
Evaluating it directly would require summing over a combinatorial number of legal symbolic states.
Instead, let $a_i$ be the active logit of slot $i$, let
$\omega_i=\exp(a_i)$, and define the elementary symmetric polynomial
$e_K(\boldsymbol\omega)=\sum_{|A|=K}\prod_{i\in A}\omega_i$,
where the sum ranges over $A\subseteq\{1,\ldots,M\}$.
After drawing $G$, the decoder samples the orbit count $K$ and active-slot set
$A$ according to
\begin{equation}
q_{\mathrm{prop}}(K\mid x_t)
\propto p_\theta(K_0=K\mid x_t)e_K(\boldsymbol\omega),\quad
q_{\mathrm{prop}}(A\mid K,x_t)
=\bigl(\prod\nolimits_{i\in A}\omega_i\bigr)/e_K(\boldsymbol\omega).
\label{eq:proposal}
\end{equation}
Here $|A|=K$. The $e_K$ factors cancel when the two proposal distributions are multiplied.
After sampling $G$, $K$,
the active set, and the selected rows, the decoder rejects proposals that violate
these constraints and draws their elements after acceptance.

\begin{proposition}
\label{prop:exact}
Conditional on acceptance, the stochastic decoder returns an exact draw from
$\widetilde p_\theta(\cdot\mid x_t)$ without evaluating its global normalizer.
\end{proposition}

Appendix~\ref{app:decoder} gives the derivation and sampling procedure.
Reverse generation repeatedly predicts a legal clean state and then applies the
same marginal forward kernel at the next noise level $t'<t$:
\begin{equation}
\hat x_0^{(t)}
\sim\widetilde p_\theta(\cdot\mid x_t),
\qquad
x_{t'}\sim q_{t'}(\cdot\mid\hat x_0^{(t)}).
\label{eq:decode-renoise}
\end{equation}
At the final sampled noise level, removing inactive slots from the clean
prediction and disregarding slot order yields the protostructure $P$.
The space group is thus resampled jointly with the other symbolic variables
throughout the reverse trajectory rather than fixed in advance.

\subsection{Symmetry-constrained geometry generation}
\label{sec:realization}

The final protostructure $P$ specifies the space group, Wyckoff occupations,
and elements.
The geometry stage generates a lattice compatible with the crystal family of
$G$ and the free fractional coordinates of one representative point, or orbit
anchor, per occupied orbit.

\paragraph{Geometry denoising.}
Building on the continuous diffusion framework of DiffCSP++ \citep{diffcsppp},
an atom-level message-passing network operates on the full conventional cell
reconstructed from the occupied orbit anchors and is conditioned on $P$, the
diffusion time, and the shared symmetry codebook. Let $b_t$ denote the noisy
lattice representation and $X_t$ the noisy full-cell fractional coordinates. The two
heads minimize
\begin{equation}
\begin{aligned}
&\mathcal L_{\mathrm{lat}}
=\mathbb E\!\left[
\left\lVert
\epsilon_G-\widehat\epsilon_\theta(b_t,X_t,P,t)
\right\rVert_2^2
\right],
\qquad \epsilon_G=\Pi_G^{\mathrm{lin}}\epsilon,\\
&\mathcal L_{\mathrm{coord}}
=\mathbb E\!\left[
\left\lVert
\Pi_{\mathcal A(P)}\widehat u_\theta(X_t,b_t,P,t)
-u_{\mathrm{wn}}^*(X_t,X_0;\sigma_t)
\right\rVert_2^2
\right].
\end{aligned}
\label{eq:geometry-objectives}
\end{equation}
Here $\epsilon\sim\mathcal N(0,I)$ is standard Gaussian lattice noise,
$\epsilon_G=\Pi_G^{\mathrm{lin}}\epsilon$ is its projection onto the
crystal-family subspace, and $\widehat\epsilon_\theta$ is the lattice-noise
prediction. The clean full-cell fractional coordinates are denoted by $X_0$,
$\sigma_t$ is the coordinate-noise scale, and $u_{\mathrm{wn}}^*$ is the
wrapped-normal denoising target in the orbit-anchor coordinates. The coordinate head predicts the denoising
field $\widehat u_\theta$. Finally, $\Pi_{\mathcal A(P)}$ maps each predicted atomic field to the free
orbit-anchor coordinates $\mathcal A(P)$ specified by $P$.

\paragraph{Symmetry-constrained sampling.}
At each reverse step, the lattice representation is projected to the subspace compatible
with the crystal family of $G$. The anchor coordinates are updated only along
their allowed degrees of freedom and expanded through the corresponding Wyckoff
operations into complete atomic orbits. By construction, every intermediate geometry
is invariant under the operations of the decoded space group $G$.
Further details on the lattice representation, geometry objectives, and reverse
sampling procedure are provided in Appendix~\ref{app:realization}.

\section{Experiments}
\label{sec:eval}
\label{sec:protocol}

We evaluate unconditional crystal generation on MP-20 \citep{cdvae} and the more challenging MPTS-52 benchmark \citep{sgequidiff}. Following standard evaluation practice \citep{symmcd}, we use each dataset's training split as the novelty reference and its test split as the coverage reference. We evaluate approximately 10k candidates per set using common relaxation, stability, novelty, and symmetry-analysis settings for comparability. The large sample sizes reduce finite-sample variation. We use publicly available generated structures for the baselines. For SGEquiDiff, these were unavailable, so we sample from its released checkpoint. Unless otherwise stated, we report five-seed means for discovery rates and structural statistics of \ours{} models and SGEquiDiff, use CHGNet \citep{chgnet} for relaxation, and identify space groups with \texttt{symprec}$=0.1$~\AA{} and an angular tolerance of $5^{\circ}$. We also independently evaluate with MACE-MP-0 \citep{mace} (Appendix~\ref{app:mace}) and perform large-scale first-principles calculations with VASP to test the robustness of \ours{}'s performance. Full evaluation settings are in Appendix~\ref{app:protocol}.

We compare against strong baselines from peer-reviewed publications, including
DiffCSP \citep{diffcsp}, MatterGen \citep{mattergen}, FlowMM \citep{flowmm},
DiffCSP++ \citep{diffcsppp}, SymmCD \citep{symmcd}, WyFormer
\citep{kazeev2025wyckoff}, and SGEquiDiff \citep{sgequidiff}.

\subsection{Symmetry and space-group distributions}
\label{sec:sg-distribution}

\begin{figure}[htb]
\centering
\begin{minipage}[t]{0.72\linewidth}
\vspace{0pt}
\centering
\includegraphics[width=\linewidth]{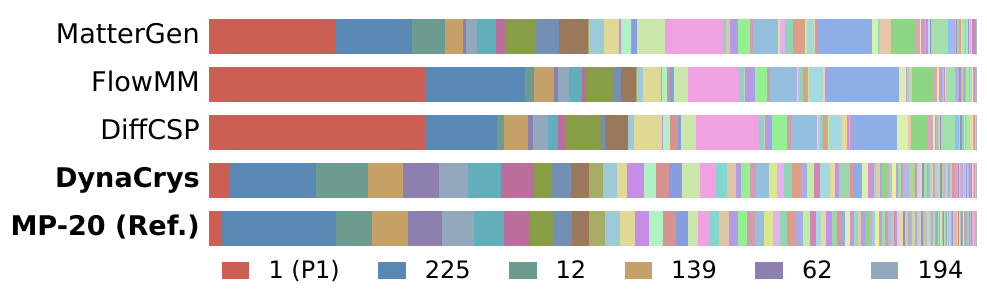}
\end{minipage}\hfill
\begin{minipage}[t]{0.265\linewidth}
\vspace{5.5973bp}
\setlength{\abovetopsep}{0pt}
\centering
\begingroup
\maintablefont
\setlength{\tabcolsep}{4pt}
\renewcommand{\arraystretch}{0.98}
\begin{tabular}{@{}lr@{}}
\toprule
Method & P1 (\%) \\
\midrule
MatterGen & 20.28 \\
FlowMM & 34.02 \\
DiffCSP & 30.79 \\
\midrule
\ours{} (ours) & 5.54 \\
\bottomrule
\end{tabular}
\endgroup

\end{minipage}
\caption{Left: space-group distributions of generated crystals on MP-20. Right: fractions of relaxed structures identified as $P1$.}
\label{fig:sg-distribution}
\end{figure}

We first assess the symmetry diversity of generated crystals. Figure~\ref{fig:sg-distribution} compares the space-group distributions of crystals generated by different models with the MP-20 reference distribution and also reports the fraction of $P1$ structures after relaxation. $P1$ has only translational symmetry and is rare among naturally occurring crystals \citep{symmcd}.

Figure~\ref{fig:sg-distribution} shows that \ours{} matches the MP-20 space-group distribution substantially more closely than the symmetry-unaware baselines. These baselines overproduce $P1$ structures, whereas \ours{} better captures the relative frequencies of a broad range of nontrivial space groups. Its $P1$ fraction also remains substantially lower after relaxation. Together, these results demonstrate a clear advantage over symmetry-unaware methods in matching the space-group statistics of known crystals and generating nontrivially symmetric structures after relaxation.

\subsection{Materials discovery}
\label{sec:results}

The stable, unique, and novel (SUN) rate \citep{mattergen} is a key metric for evaluating crystal generation. SUN@$\tau$ is the fraction of generated candidates that satisfy the stability criterion $E_{\mathrm{hull}}\leq\tau$ ($\tau=0$ and $0.1$~eV/atom), are unique within the generated set, and are novel relative to the training set. SSUN@$\tau$ additionally requires a non-$P1$ space group after relaxation \citep{kazeev2025wyckoff}.

We report results for \ours{} and \ours{}-II, two models that share the same framework and differ only in their choices of optional parameter settings (Appendix~\ref{app:den-film-stage}). Tables~\ref{tab:mp20-discovery} and~\ref{tab:mpts52-discovery} compare their discovery performance with the baselines. We also conduct large-scale first-principles validation on MP-20, evaluating 2,000 candidates from each model's unique-and-novel pool with VASP at a stability threshold of 0.1~eV/atom (Appendix~\ref{app:vasp}). By the standards of crystal-generation research, this is a particularly large-scale first-principles evaluation.

\begin{table}[htb]
\centering
\caption{MP-20 material discovery rates (\%; higher is better). \textbf{Bold} highlights leading results; underlining marks runner-up results. The same convention applies to subsequent result tables.}
\label{tab:main}
\label{tab:mp20-discovery}
\begingroup
\small
\setlength{\tabcolsep}{3pt}
\renewcommand{\arraystretch}{1.06}
\begin{tabular*}{\linewidth}{@{\extracolsep{\fill}}lcccccc@{}}
\toprule
Method & SUN@0 & SSUN@0 & SUN@0.1 & SSUN@0.1 & SUN-VASP & SSUN-VASP \\
\midrule
DiffCSP++ & 7.71 & 7.64 & 37.95 & 37.31 & 24.47 & 24.47 \\
SymmCD & 7.34 & 7.18 & 33.11 & 32.02 & 21.56 & 21.52 \\
WyFormer & 6.76 & 6.69 & 33.28 & 32.11 & 19.93 & 19.75 \\
SGEquiDiff & 8.45 & 8.31 & 38.00 & 36.38 & 26.74 & 26.62 \\
\midrule
\ours{} (ours) & \textbf{9.39} & \textbf{9.31} & \textbf{41.49} & \textbf{40.12} & \textbf{29.56} & \textbf{28.98} \\
\ours{}-II (ours) & \textbf{9.18} & \textbf{9.10} & \textbf{41.32} & \textbf{39.87} & \textbf{30.63} & \textbf{30.28} \\
\bottomrule
\end{tabular*}
\endgroup

\end{table}

\Needspace{120pt}
\begin{wraptable}[11]{r}{248pt}
\raisebox{9pt}{%
\begin{minipage}[t]{\linewidth}
\setlength{\abovecaptionskip}{0pt}
\centering
\caption{MPTS-52 material discovery rates (\%; higher is better).}
\label{tab:mpts52-discovery}
\begingroup%
\small%
\setlength{\tabcolsep}{4pt}%
\renewcommand{\arraystretch}{1.08}%
\begin{tabular}{@{}lrrrr@{}}
\toprule
Method & SUN@0 & SSUN@0 & SUN@0.1 & SSUN@0.1 \\
\midrule
SymmCD & 5.07 & 4.88 & 21.65 & 20.64 \\
WyFormer & 2.05 & 1.81 & 13.58 & 10.57 \\
SGEquiDiff & 5.17 & 4.94 & 26.29 & 24.33 \\
\midrule
\ours{} (ours) & \textbf{7.30} & \textbf{7.16} & \textbf{34.20} & \textbf{33.05} \\
\ours{}-II (ours) & \textbf{6.65} & \textbf{6.58} & \textbf{31.85} & \textbf{30.85} \\
\bottomrule
\end{tabular}%
\endgroup%

\end{minipage}%
}
\end{wraptable}

\ours{} achieves the best SUN and SSUN rates at both thresholds on both datasets (Tables~\ref{tab:mp20-discovery} and~\ref{tab:mpts52-discovery}). Its lead over the baselines becomes more pronounced on the more challenging MPTS-52 dataset. VASP (Table~\ref{tab:mp20-discovery}) and MACE-MP-0 evaluation (Appendix~\ref{app:mace}) further separately confirm this lead. These results demonstrate \ours{}'s leading and robust discovery performance both overall and under the additional requirement of nontrivial symmetry after relaxation.

\subsection{Validity and distributional fidelity}
\label{sec:quality}

We also assess structural and compositional validity, distributional distances for density, the number of distinct elements, and Wyckoff degrees of freedom, and coverage recall and precision on relaxed structures \citep{sgequidiff}. These metrics measure structural and chemical plausibility and agreement with the reference data. The median energy above the hull ($E_{\mathrm{hull}}$) characterizes energetic quality (Tables~\ref{tab:mp20-quality} and~\ref{tab:mpts52-quality}). We compute distributional distances on valid structures, using the test split for MP-20, which is randomly partitioned. For MPTS-52, whose chronological split produces a substantial train--test distribution shift (Appendix~\ref{app:quality-reference}), we use the training split as the reference to assess how well generated structures capture the statistical properties of the training materials.

\begin{table}[htb]
\centering
\caption{Results for validity, distributional discrepancies, and energetic properties on MP-20.}
\label{tab:quality}
\label{tab:mp20-quality}
\begingroup
\maintablefont
\setlength{\tabcolsep}{2pt}
\renewcommand{\arraystretch}{1.08}
\begin{tabular*}{\linewidth}{@{\extracolsep{\fill}}lccccccccc@{}}
\toprule
Method & \multicolumn{2}{c}{Validity (\%) $\uparrow$} & \multicolumn{3}{c}{Distribution $\downarrow$} & \multicolumn{2}{c}{Coverage (\%) $\uparrow$} & P1 (\%) $\downarrow$ & Median $E_{\mathrm{hull}}$ $\downarrow$ \\
\cmidrule(lr){2-3}\cmidrule(lr){4-6}\cmidrule(lr){7-8}
 & Struct. & Comp. & $d_\rho$ & $d_{N_{\mathrm{el}}}$ & $d_{\mathrm{Wyckoff}}$ & R & P & & (eV/atom) \\
\midrule
DiffCSP++ & 99.91 & 85.12 & \textemdash & \textemdash & \textemdash & \textemdash & \textemdash & \textbf{5.10} & 0.107 \\
SymmCD & 98.54 & \textbf{86.27} & 0.2363 & 0.3902 & 0.0722 & \underline{99.78} & 99.24 & 7.60 & 0.115 \\
WyFormer & 99.84 & 82.65 & 0.2081 & 0.0753 & 0.0590 & \textbf{99.79} & 99.00 & 6.43 & 0.130 \\
SGEquiDiff & 99.50 & \underline{86.13} & 0.1761 & 0.1195 & 0.0781 & 99.76 & 99.14 & 7.95 & 0.082 \\
\midrule
\ours{} (ours) & \textbf{99.97} & 83.36 & \textbf{0.1025} & \textbf{0.0518} & \textbf{0.0340} & 99.71 & \textbf{99.31} & 5.54 & \textbf{0.072} \\
\ours{}-II (ours) & \textbf{99.95} & 83.30 & \textbf{0.1225} & \textbf{0.0627} & \textbf{0.0344} & 99.70 & \textbf{99.42} & \underline{5.42} & \textbf{0.060} \\
\bottomrule
\end{tabular*}
\endgroup

\end{table}

\begin{table}[htb]
\centering
\caption{Results for validity, distributional discrepancies, and energetic properties on MPTS-52.}
\label{tab:mpts52-quality}
\begingroup
\maintablefont
\setlength{\tabcolsep}{2pt}
\renewcommand{\arraystretch}{1.08}
\begin{tabular*}{\linewidth}{@{\extracolsep{\fill}}lccccccccc@{}}
\toprule
Method & \multicolumn{2}{c}{Validity (\%) $\uparrow$} & \multicolumn{3}{c}{Distribution $\downarrow$} & \multicolumn{2}{c}{Coverage (\%) $\uparrow$} & P1 (\%) $\downarrow$ & Median $E_{\mathrm{hull}}$ $\downarrow$ \\
\cmidrule(lr){2-3}\cmidrule(lr){4-6}\cmidrule(lr){7-8}
 & Struct. & Comp. & $d_\rho$ & $d_{N_{\mathrm{el}}}$ & $d_{\mathrm{Wyckoff}}$ & R & P & & (eV/atom) \\
\midrule
SymmCD & 89.25 & 73.53 & \textbf{0.1372} & 0.5610 & 0.1054 & 99.78 & 89.78 & 9.37 & 0.204 \\
WyFormer & 90.62 & 70.59 & 0.6904 & 0.2382 & 0.2201 & \textbf{99.96} & 90.56 & 23.80 & 0.323 \\
SGEquiDiff & 98.36 & \textbf{79.91} & 0.2400 & 0.1479 & 0.1266 & \underline{99.94} & 96.24 & 10.09 & 0.161 \\
\midrule
\ours{} (ours) & \textbf{99.91} & 75.02 & 0.2618 & \textbf{0.0501} & \textbf{0.0626} & 99.83 & \textbf{98.54} & \textbf{4.59} & \textbf{0.093} \\
\ours{}-II (ours) & \textbf{99.90} & \underline{76.90} & \underline{0.1779} & \textbf{0.0202} & \textbf{0.0501} & 99.84 & \textbf{98.35} & \textbf{4.66} & \textbf{0.101} \\
\bottomrule
\end{tabular*}
\endgroup

\end{table}

\ours{} leads in structural validity and coverage precision on both datasets, while also achieving the smallest element-count and Wyckoff distribution distances and the lowest median energies above the hull (Tables~\ref{tab:mp20-quality} and~\ref{tab:mpts52-quality}). On MP-20, it also leads the density-distribution comparison. These results show that \ours{} combines the highest SUN and SSUN rates with leading structural validity, favorable energies above the hull, and close agreement with key statistics of the reference materials.

\begin{table}[htb]
\centering
\begin{minipage}[t]{195.86282pt}
\vspace{0pt}
\centering
\captionof{table}{Geometric changes under relaxation.}
\label{tab:rmsd}
\begingroup%
\maintablefont%
\setlength{\tabcolsep}{1pt}%
\renewcommand{\arraystretch}{1.08}%
\begin{tabular}{@{}lcccc@{}}
\toprule
 & \multicolumn{2}{c}{MP-20} & \multicolumn{2}{c}{MPTS-52} \\
\cmidrule(lr){2-3}\cmidrule(lr){4-5}
 & RMSD $\downarrow$ & Matched $\uparrow$ & RMSD $\downarrow$ & Matched $\uparrow$ \\
Method & (\AA) & (\%) & (\AA) & (\%) \\
\midrule
DiffCSP++ & 0.206 & 95.29 & --- & --- \\
SymmCD & 0.377 & 85.27 & 0.523 & 75.64 \\
WyFormer & 0.191 & 95.54 & 0.641 & 45.60 \\
SGEquiDiff & 0.190 & 91.38 & 0.294 & 88.05 \\
\midrule
\ours{} (ours) & \textbf{0.172} & \textbf{96.18} & \textbf{0.255} & \textbf{97.11} \\
\ours{}-II (ours) & \textbf{0.164} & \textbf{96.65} & \textbf{0.239} & \textbf{96.95} \\
\bottomrule
\end{tabular}%
\endgroup%

\end{minipage}\hfill
\begin{minipage}[t]{195.62217pt}
\vspace{7.875pt}
\centering
\makebox[180.37pt][l]{\hspace*{8.197087113pt}\raisebox{1.968323413pt}[114.74434pt][0pt]{\includegraphics[width=170.034585978pt]{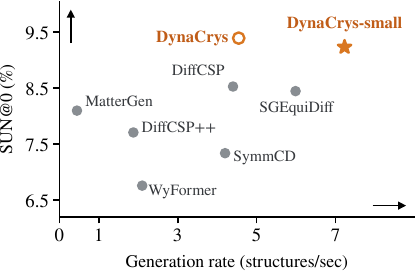}}}
\captionof{figure}{Generation rates of crystals on MP-20.}
\label{fig:efficiency}
\end{minipage}
\end{table}

\subsection{Geometric quality and computational efficiency}
\label{sec:rmsd}
\label{sec:efficiency}

We assess geometric quality by comparing generated structures with their relaxed endpoints. We report the mean root-mean-square displacement (RMSD) over matched pairs and the fraction of evaluated pairs with a structural correspondence (Matched), using MatterGen's matching protocol. \ours{} achieves the lowest RMSD and highest matching rates on both datasets (Table~\ref{tab:rmsd}). These results show that \ours{} generates higher-quality geometries that require smaller adjustments during relaxation and more often retain structural correspondence with their relaxed endpoints.

We also evaluate generation efficiency and the effect of fewer geometry sampling steps on MP-20. \ours{} achieves the highest SUN@0 among the compared methods, while \ourssmall{} (trained within the same framework with a reduced hidden dimension; Appendix~\ref{app:realization}) achieves a similar SUN rate and the highest generation rate for complete structures (Figure~\ref{fig:efficiency}; timing settings in Appendix~\ref{app:protocol-timing}). Besides, we find that halving \ours{}'s geometry sampling steps from 1,000 to 500 preserves approximately 93.8\% of its SUN@0.1 rate (full results in Appendix~\ref{app:geometry-steps}).

\Needspace{5\baselineskip}
\subsection{Ablation studies}
\label{sec:ablations}

To assess the contributions of individual design choices, we conduct the following ablation studies on MPTS-52. Table~\ref{tab:core-ablation} reports their discovery and stability results. All ablations show worse SUN rates. A detailed comparison including additional metrics is provided in Appendix~\ref{app:full-ablation} (Table~\ref{tab:full-ablation}).

\emph{Dynamic space-group generation.} We train a variant without dynamic space-group diffusion: it samples the space group once and keeps it fixed throughout generation. Fixing the space group reduces SUN@0.1 by 4.77 percentage points, lowers SSUN and stability, and increases median $E_{\mathrm{hull}}$. These results demonstrate the benefit of jointly revising the space group, occupations, and elements for discovering stable new crystals.

\emph{Group--subgroup diffusion kernel.} We retain dynamic space-group generation and replace the group--subgroup kernel with an all-to-all kernel defined on a fully connected graph. The all-to-all variant better matches the reference Wyckoff distribution, but matches the element-count and density distributions less closely and yields lower SUN rates.

\begin{table}[htb]
\centering
\caption{Ablation studies on MPTS-52, averaged over five generation seeds. The indented $\Delta$ row is relative to w/o FiLM. Rates are in \% (changes in percentage points) and energies in eV/atom.}
\label{tab:core-ablation}
\label{tab:decoder-ablation}
\begingroup
\maintablefont
\setlength{\tabcolsep}{1pt}
\renewcommand{\arraystretch}{1.12}
\begin{tabular*}{\linewidth}{@{\extracolsep{\fill}}lcccccc@{}}
\toprule
& \multicolumn{4}{c}{Discovery (\%) $\uparrow$} & \multicolumn{2}{c}{Stability} \\
\cmidrule(lr){2-5}\cmidrule(lr){6-7}
Method & SUN@0 & SSUN@0 & SUN@0.1 & SSUN@0.1 & Stable@0.1 $\uparrow$ & Median $E_{\mathrm{hull}}$ $\downarrow$ \\
\midrule
\textcolor[HTML]{C05C17}{\textbf{\ours{} (ours)}} & \textbf{7.30} & \textbf{7.16} & \textbf{34.20} & \textbf{33.05} & \textbf{51.87} & \textbf{0.093} \\
\midrule
\textbf{Ablation: w/o Dynamic Space Group} & 6.32 & 6.21 & 29.43 & 28.32 & 46.00 & 0.118 \\
\midrule
\textbf{Ablation: w/o Group-subgroup Graph} & 6.71 & 6.64 & 31.61 & 30.90 & 50.68 & 0.097 \\
\midrule
\textbf{Ablation: w/o FiLM} & 6.51 & 6.41 & 32.91 & 31.73 & 49.29 & 0.102 \\
\hspace{0.6em}$\hookrightarrow$\,\textbf{Ablation: w/o stochastic decoding ($\Delta$)} & \strut{\fontsize{7.5pt}{10pt}\selectfont(-1.16)} & \strut{\fontsize{7.5pt}{10pt}\selectfont(-1.13)} & \strut{\fontsize{7.5pt}{10pt}\selectfont(-5.26)} & \strut{\fontsize{7.5pt}{10pt}\selectfont(-4.67)} & \strut{\fontsize{7.5pt}{10pt}\selectfont(-2.56)} & \strut{\fontsize{7.5pt}{10pt}\selectfont(+0.012)} \\
\bottomrule
\end{tabular*}
\endgroup

\end{table}

\emph{Element-head fine-tuning.} We omit the element-head fine-tuning stage during training to obtain the w/o FiLM variant. The results show that removing this stage improves agreement with the reference element-count distribution, but worsens agreement with the density and Wyckoff distributions and lowers stability.

\emph{Stochastic legal decoding.} Using the same checkpoint before element-head fine-tuning and the same geometry model, we compare stochastic constrained sampling with maximum a posteriori (MAP) decoding, which deterministically searches for high-probability legal configurations. MAP achieves higher validity and lower RMSD, but shows poorer agreement with the reference element-count distribution and lower stability and SUN rates.

\section{Conclusion}
\label{sec:conclusion}

We introduced \ours{}, a framework for complete crystal generation in which the space group jointly evolves and is dynamically revised with Wyckoff occupations and elements. A group--subgroup diffusion kernel guides space-group updates, while a legality-constrained stochastic decoder produces crystallographically compatible protostructures for subsequent symmetry-constrained geometry generation. \ours{} achieves state-of-the-art performance in symmetry-aware crystal generation on standard benchmarks, while supporting fast sampling. Independent MLIP evaluations and large-scale first-principles validation confirm this advantage, and ablations show the contributions of the key design choices. \ours{} thus offers a new, strong, and more flexible framework for crystal generation. A limitation of our work is that the experimental synthesizability of the generated crystals remains unverified, as our stability assessments rely on computational screening proxies, including hull distances computed against a finite snapshot of reference structures.

\label{end:main-text}

\clearpage
\appendix
\section{Symbolic State Space and Legality}
\label{app:legalset}

This appendix defines the finite state space of the symbolic model and the
legal clean states used by the decoder.

\subsection{Symbolic state space}
\label{app:paddedset}

Let
\[
\mathcal G=\{G:n_G^{\mathrm{train}}>0\}
\]
be the dataset-specific vocabulary of training-supported space groups: 168 for
MP-20 and 202 for MPTS-52. Using PyXtal's
space group and Wyckoff position tables \citep{pyxtal}, we construct the
crystallographic row catalogue and associated symmetry operations for all 230
space groups, with \(\mathcal R_G\) denoting the valid Wyckoff rows of group
\(G\). Across this catalogue there are 1,731 valid \((G,r)\) pairs, and the
row table for each space group is padded to width 27.

The symbolic state uses \(M=20\) orbit slots for MP-20 and \(M=52\) for
MPTS-52. A clean state has
\[
G\in\mathcal G,\qquad
K\in\{1,\ldots,M\},\qquad
\sum_{i=1}^{M}m_i=K.
\]
Each active slot has \(r_i\in\mathcal R_G\) and carries a categorical element
label \(Z_i\in\mathcal Z\), while inactive slots use separate PAD symbols.
Noisy states replace the row index by the continuous row vector \(y_i\) defined
in Appendix~\ref{app:corruption}.

\subsection{Crystallographic tables}
\label{app:rowvocab}

The symbolic and geometry stages share the following crystallographic metadata
for each valid \((G,r)\):
\begingroup
\renewcommand{\arraystretch}{1.10}
\begin{center}
\small
\begin{tabular}{@{}cl@{}}
\toprule
\textbf{Symbol} & \textbf{Meaning} \\
\midrule
\(\mu_{G,r}\) & multiplicity in the conventional cell \\
\(\operatorname{dof}_{G,r}\) & number of free anchor coordinates \\
\(s_{G,r}\) & site-symmetry label \\
\(e_{G,r}\) & enumeration within \((G,s_{G,r})\) \\
\bottomrule
\end{tabular}
\end{center}
\endgroup
The valid-row mask restricts every space-group-conditional row distribution to
\(\mathcal R_G\). Multiplicity enters the atom-count constraint, while the
remaining metadata parameterize the shared symmetry codebook and the geometry
model. The configuration-specific capacity \(u_{G,r}\) is defined below.

\subsection{Capacity and legal clean states}
\label{app:capacity}
\label{app:legalmap}

For MP-20 configuration \ours{}, let
\(q_{0.99}^{\mathrm{train}}(G,r)\) be the 99th percentile of the
within-structure repetition count of row \((G,r)\), computed on the training
partition. Let
\(B_{G,r}=\lfloor N_{\max}/\mu_{G,r}\rfloor\) be the largest occupation
compatible with the conventional-cell atom cap \(N_{\max}\), set to 80 for
MP-20 and 208 for MPTS-52. The resulting empirical capacity is
\begin{equation}
u_{G,r}=
\begin{cases}
0,
& r\notin\mathcal R_G,\\
1,
& \operatorname{dof}_{G,r}=0
  \text{ or }q_{0.99}^{\mathrm{train}}(G,r)\text{ is unavailable},\\
\displaystyle
\max\!\left\{
1,\,
\min\!\left(
\left\lceil q_{0.99}^{\mathrm{train}}(G,r)\right\rceil,\,
B_{G,r}
\right)
\right\},
& \text{otherwise}.
\end{cases}
\label{eq:capacity-rule}
\end{equation}
MP-20 configuration \ours{}-II and both MPTS-52 configurations use the
multiplicity-based capacity
rule in Appendix~\ref{app:den-film-stage}, which also records the
configuration differences.

Write a padded clean state as \(x_0=(G,K,A,R,Z)\), where
\(A\subseteq\{1,\ldots,M\}\) is the active-slot set and
\(R=(r_i)_{i\in A}\), \(Z=(Z_i)_{i\in A}\). Define the number of active
occupations assigned to row \(r\) by
\[
n_r(A,R)=\sum_{i\in A}\mathbf 1[r_i=r].
\]
For a chosen capacity table \(u\), the legal clean-state set in
Eq.~\eqref{eq:target} is
\begin{equation}
\mathcal X_{\mathrm{legal}}\equiv\mathcal X_{\mathrm{legal}}(u)=
\left\{
\begin{aligned}
&(G,K,A,R,Z):\quad G\in\mathcal G,\quad |A|=K,\\
&(r_i,Z_i)\in\mathcal R_G\times\mathcal Z\quad \forall i\in A,\\
&n_r(A,R)\le u_{G,r}\quad \forall r\in\mathcal R_G,\\
&\displaystyle\sum_{i\in A}\mu_{G,r_i}\le N_{\max}
\end{aligned}
\right\}.
\label{eq:legal-set}
\end{equation}

\section{Forward Corruption Kernels}
\label{app:corruption}

The symbolic forward process is a family of marginal kernels
\(q_t(x_t\mid x_0)\). At each noise level it samples
\[
G_t\longrightarrow K_t\longrightarrow
\text{survivor--birth--PAD partition}
\longrightarrow \{(m_{i,t},Z_{i,t},y_{i,t})\}_{i=1}^{M}.
\]
The same family is used for training-time corruption and reverse-time
re-noising.

\subsection{Training priors}
\label{app:corr-priors}

All reset and birth distributions are estimated on the training partition.
Writing \(n_G\), \(n_{G,r}\), and \(n_Z\) for the corresponding counts,
\begin{equation}
\pi_G(G)=\frac{n_G}{\sum_{G'\in\mathcal G}n_{G'}},
\qquad
\pi_Z(Z)=\frac{n_Z}{\sum_{Z'\in\mathcal Z}n_{Z'}}.
\label{eq:priors-gz}
\end{equation}
The row prior is smoothed within the valid vocabulary of each group:
\begin{equation}
\pi_R(r\mid G)
=\frac{n_{G,r}+1}
{\sum_{r'\in\mathcal R_G}(n_{G,r'}+1)},
\qquad r\in\mathcal R_G.
\label{eq:prior-row}
\end{equation}
For the orbit count, let
\(\widehat p_{\mathrm{train}}(K\mid G)\) and
\(\widehat p_{\mathrm{train}}(K)\) be the space-group-conditional and global
empirical distributions. We use
\begin{equation}
\pi_K(K\mid G)
=(1-\rho_G)\widehat p_{\mathrm{train}}(K\mid G)
+\rho_G\widehat p_{\mathrm{train}}(K),
\qquad
\rho_G=\frac{\alpha_K}{n_G+\alpha_K},
\quad \alpha_K=20.
\label{eq:prior-count}
\end{equation}
Here \(\alpha_K=20\) is a smoothing pseudo-count controlling shrinkage toward
the global orbit-count distribution.

\subsection{Space-group graph and kernel}
\label{app:corr-graph}
\label{app:corr-kernel}

The graph on \(\mathcal G\) contains the maximal
translationengleiche and klassengleiche group--subgroup
relations, with edge directions removed and self-relations omitted. Let \(A\)
be its adjacency matrix,
\(\widetilde A=A+I\), \(D=\operatorname{diag}(\widetilde A\mathbf 1)\), and
\begin{equation}
W_{\mathcal G}=D^{-1}\widetilde A,\qquad
Q=W_{\mathcal G}-I,\qquad
R_s=\exp\!\left(\beta_{\max}sQ\right),
\quad \beta_{\max}=4,
\label{eq:ctmc}
\end{equation}
where \(s=t/T\). The space-group channel is
\begin{equation}
q_t^G(G_t\mid G_0)
=\lambda_G(s)R_s(G_t\mid G_0)
+\bigl[1-\lambda_G(s)\bigr]\pi_G(G_t).
\label{eq:g-kernel-full}
\end{equation}

\subsection{Orbit-count and slot channels}
\label{app:corr-channels}

Given the sampled group, the orbit count follows
\begin{equation}
q_t^K(K_t\mid K_0,G_t)
=\lambda_K(s)\delta_{K_0}(K_t)
+\bigl[1-\lambda_K(s)\bigr]\pi_K(K_t\mid G_t).
\label{eq:kkernel}
\end{equation}
The number of surviving clean occupations is
\begin{equation}
S_t\sim
\operatorname{Binomial}\!\left(
\min(K_0,K_t),\,\alpha_S(s)
\right).
\label{eq:survivors}
\end{equation}
The \(S_t\) survivors are matched to distinct clean occupations sampled
uniformly without replacement. Together with \(K_t-S_t\) births, they are
placed uniformly among the \(M\) slots; the remaining slots are PAD.

Let \(c_{G,r}\) and \(c_{\mathrm{pad}}\) be the row and padding vectors of
the shared codebook in Appendix~\ref{app:codebook}. For each slot,
\begin{equation}
y_{i,t}=\mu_{i,t}+\sigma_y(s)\epsilon_i,
\qquad \epsilon_i\sim\mathcal N(0,I_d),
\label{eq:row-channel}
\end{equation}
with
\begin{equation}
\mu_{i,t}=
\begin{cases}
\sqrt{\bar\alpha_y(s)}\,c_{G_0,r_i^0},
& \text{survivor matched to }r_i^0,\\
c_{G_t,r_i^{\mathrm{birth}}},
\quad r_i^{\mathrm{birth}}\sim\pi_R(\cdot\mid G_t),
& \text{birth},\\
c_{\mathrm{pad}},
& \text{PAD}.
\end{cases}
\label{eq:ychannel}
\end{equation}
At survivor slots, elements follow
\begin{equation}
q_t^Z(Z_{i,t}\mid Z_i^0)
=\alpha_Z(s)\delta_{Z_i^0}(Z_{i,t})
+\bigl[1-\alpha_Z(s)\bigr]\pi_Z(Z_{i,t}).
\label{eq:element-channel}
\end{equation}
Birth elements are drawn from \(\pi_Z\), while PAD slots carry the element
PAD symbol.

\subsection{Schedules and terminal prior}
\label{app:corr-schedules}
\label{app:corr-endpoint}

The discrete channels share a quadratic retention schedule, and the row
channel uses a cosine signal schedule:
\begin{equation}
\begin{aligned}
\lambda_G(s)=\lambda_K(s)=\alpha_S(s)=\alpha_Z(s)=(1-s)^2,
\\
\bar\alpha_y(s)=\cos^2\!\left(\frac{\pi s}{2}\right),
\qquad
\sigma_y(s)=\sqrt{1-\bar\alpha_y(s)}.
\end{aligned}
\label{eq:schedules}
\end{equation}
Here \(\bar\alpha_y(s)\) is the retained row-signal power and \(\sigma_y(s)\)
is the row-noise scale. The training horizon is \(T=1000\); at generation
time, the reverse sampler operates at 200 noise levels arranged in descending
order along this schedule.

At \(s=1\), all clean-state retention terms vanish. The terminal state can
therefore be sampled directly as
\begin{equation}
\begin{aligned}
&G_T\sim\pi_G,\qquad
K_T\sim\pi_K(\cdot\mid G_T),\qquad
A_T\sim\operatorname{Unif}\{A\subseteq\{1,\ldots,M\}:|A|=K_T\},\\
&r_{i,T}\sim\pi_R(\cdot\mid G_T),\qquad
Z_{i,T}\sim\pi_Z,\qquad
y_{i,T}=c_{G_T,r_{i,T}}+\epsilon_i
\quad (i\in A_T).
\end{aligned}
\label{eq:terminal-prior}
\end{equation}
Here \(m_{i,T}=\mathbf 1[i\in A_T]\). On PAD slots,
\(y_{i,T}=c_{\mathrm{pad}}+\epsilon_i\), and \(Z_{i,T}\) is the element PAD
symbol. This distribution contains no dependence on \(x_0\) and is the
starting distribution of the reverse sampler.

\section{Shared Symmetry Codebook}
\label{app:codebook}

The codebook maps space groups and space-group-specific Wyckoff rows into a common
vector space. It is constructed once, frozen, and shared by the symbolic
and geometry stages.

\subsection{Space-group and row representations}
\label{app:cb-encoder}

For a symmetry operation \(g=(R\mid\tau)\) in the fractional-coordinate
basis, define
\begin{equation}
\phi(g)=
\left[
\operatorname{vec}(R),\,
\tau\bmod 1,\,
\det R,\,
\operatorname{tr}R
\right]\in\mathbb R^{14}.
\label{eq:operation-features}
\end{equation}
A shared operation encoder \(e_\phi\) maps these features to
\(\mathbb R^d\). Let \(S_G\) denote the operation set of space group \(G\).
A learned map \(f_G\) produces the space-group embedding from the pooled
operation features:
\begin{equation}
c_G=f_G\!\left[
\operatorname{mean}_{g\in S_G}e_\phi(g)
\;\middle\Vert\;
\operatorname{max}_{g\in S_G}e_\phi(g)
\right].
\label{eq:cbgroup}
\end{equation}

For row \(r\), let \(\mathcal C_{G,r}\) be its orbit-generating coset
representatives and \(\mathcal H_{G,r}\) the stabilizer of its
representative. A learned map \(f_R\) produces the row embedding from the
space-group embedding, operation-derived features, and row metadata:
\begin{equation}
\begin{aligned}
c_{G,r}=f_R\Big[
&c_G,\,
\operatorname{mean}_{g\in\mathcal C_{G,r}}e_\phi(g),\,
\operatorname{mean}_{g\in\mathcal H_{G,r}}e_\phi(g),\\
&\operatorname{emb}(\mu_{G,r}),\,
\operatorname{emb}(\operatorname{dof}_{G,r}),\,
\operatorname{emb}(s_{G,r}),\,
\operatorname{emb}(e_{G,r})
\Big].
\end{aligned}
\label{eq:cbrow}
\end{equation}
Here \(e_{G,r}\) enumerates rows within the same
\((G,\text{site-symmetry})\) class. The codebook width is \(d=128\), and
each metadata embedding has width 16. A separately initialized
\(c_{\mathrm{pad}}\) represents padded rows.

\subsection{Pretraining and freezing}
\label{app:cb-stage1}
\label{app:cb-stage2}
\label{app:cb-bake}

Pretraining has two stages. The first fits deterministic crystallographic
labels from operation-derived representations. From \(c_G\), the model
predicts the space-group type, crystal system, centering, and point group;
an operation-only row representation predicts the number of free
coordinates and the site-symmetry label. The sum of these cross-entropies
is denoted by \(\mathcal L_{\mathrm{struct}}\).

The second stage freezes the symmetry-operation encoder and the space-group encoder and refines
the modules that construct the row embeddings. A row center is corrupted with the row-channel
cosine schedule in Eq.~\eqref{eq:schedules},
\[
\widetilde c_{G,r}
=\sqrt{\bar\alpha_y(s)}\,c_{G,r}
+\sqrt{1-\bar\alpha_y(s)}\,\epsilon,
\qquad \epsilon\sim\mathcal N(0,I_d),
\]
and retrieved among the valid rows of the same group with a cosine-softmax
loss
\[
\mathcal L_{\mathrm{retr}}
=-\log
\frac{
\exp\!\left(\cos(\widetilde c_{G,r},c_{G,r})/\tau_{\mathrm{retr}}\right)
}{
\displaystyle\sum_{r'\in\mathcal R_G}
\exp\!\left(\cos(\widetilde c_{G,r},c_{G,r'})/\tau_{\mathrm{retr}}\right)
},
\]
where \(\tau_{\mathrm{retr}}\) is the retrieval temperature. Distinct rows
within a group are separated by
\begin{equation}
\mathcal L_{\mathrm{sep}}
=\underset{\substack{G,\ r\ne r'\\r,r'\in\mathcal R_G}}
{\operatorname{mean}}
\left[
\operatorname{ReLU}\!\left(
\cos(c_{G,r},c_{G,r'})-m
\right)^2
\right].
\label{eq:codebook-separation}
\end{equation}
The second-stage objective is
\begin{equation}
\mathcal L_{\mathrm{cb}}
=\mathcal L_{\mathrm{retr}}
+\lambda_{\mathrm{sep}}\mathcal L_{\mathrm{sep}}.
\label{eq:codebook-objective}
\end{equation}

The shared training settings are summarized below.
\begin{center}
\small
\begin{tabular}{@{}ll@{}}
\toprule
\textbf{Setting} & \textbf{Value} \\
\midrule
operation / codebook width & 128 / 128 \\
metadata width & 16 \\
steps per stage & 3,000 \\
optimizer & AdamW~\citep{adamw} with cosine decay \\
learning rate & \(3\times10^{-3}\) \\
stage-1 / stage-2 weight decay & 0 / 0.01 \\
retrieval temperature / noise draws & 0.1 / 4 \\
separation margin \(m\) / weight \(\lambda_{\mathrm{sep}}\) & 0.5 / 1 \\
\bottomrule
\end{tabular}
\end{center}

After training, every nonzero vector is normalized as
\begin{equation}
c\leftarrow\sqrt d\,\frac{c}{\lVert c\rVert_2},
\label{eq:codebook-normalization}
\end{equation}
so each vector has unit root-mean-square coordinate magnitude. The
resulting \(c_G\), \(c_{G,r}\), and \(c_{\mathrm{pad}}\) tables are frozen.

\subsection{Use in \ours{}}
\label{app:cb-roles}

The frozen \(c_{G_t}\) represents the current noisy space group in the symbolic
denoiser. The row vectors \(c_{G,r}\) serve both as centers of the
continuous row channel and as the tied vectors of the space-group-conditional
row head. The accepted \(c_G\) and \(c_{G,r}\) also condition the geometry
model. All reported \ours{} configurations use the same frozen codebook.

\section{Symbolic Denoiser and Training}
\label{app:denoiser}

This appendix specifies the set denoiser, its prediction heads, and the
objective summarized in Section~\ref{sec:denoiser}.

\subsection{Permutation-equivariant set denoiser}
\label{app:den-arch}

The noisy state is encoded as one global token and $M$ orbit-slot tokens:
\begin{equation}
h_{\mathrm{global}}^{\mathrm{in}}
=\mathrm{MLP}\!\left[c_{G_t},e_{K_t},e_t\right],
\qquad
h_i^{\mathrm{in}}
=\mathrm{MLP}\!\left[e_{m_i},e_{Z_i},y_i,e_t,c_{G_t}\right].
\label{eq:tokens}
\end{equation}
Here $e_t$ is a sinusoidal time embedding passed through an MLP. A Transformer
encoder processes the tokens without slot positional encodings. Its slot
outputs are therefore equivariant to permutations of the orbit occupations,
while the global output is invariant to those permutations.

Linear heads on the global output predict $G_0$ and $K_0$. A slot-wise active
head predicts whether each output corresponds to an occupied orbit or padding.
The remaining slot heads predict its Wyckoff row and element.

\subsection{Row and element heads}
\label{app:den-heads}

For a candidate space group $G$, the row head classifies only over
$\mathcal R_G$. Its logits are tied to the shared codebook with RMS
normalization \citep{rmsnorm}:
\begin{equation}
\begin{aligned}
&\ell_{i,r}^{(G)}
=\frac{\widehat q_i^{\top}\widehat c_{G,r}}{\tau}+b_{G,r},\\
&\widehat q_i
=\operatorname{normalize}\!\left(
\operatorname{RMSNorm}(W_qh_i)
+g_y(t)\lambda_y\operatorname{RMSNorm}(y_i)
\right),
\end{aligned}
\label{eq:rowhead}
\end{equation}
where $\widehat c_{G,r}=\operatorname{normalize}(c_{G,r})$, $\tau$ and
$\lambda_y$ are learned scalars, $b_{G,r}$ is a row bias, $s=t/T$, and
$g_y(t)=\sqrt{\bar\alpha_y(s)}$. The invalid-row logits are masked before the
row-wise softmax. During training, the clean space group $G_0$ determines the
row vocabulary $\mathcal R_{G_0}$; during sampling, the row head is evaluated
over $\mathcal R_G$ for each proposed clean space group $G$.

The element head conditions its logits on the selected $(G,r)$ through
feature-wise linear modulation (FiLM) \citep{film}:
\begin{equation}
\ell_i^Z(G,r)
=\left[1+\gamma_{\mathrm{raw}}(c_{G,r})\right]\odot W_Zh_i
+\beta(c_{G,r}),
\label{eq:film}
\end{equation}
where a shared MLP produces $\gamma_{\mathrm{raw}}$ and $\beta$ from the row
code. The modulation is initialized at the identity. Its regularizer is
\begin{equation}
\mathcal L_{\mathrm{FiLM}}
=\lambda_{\mathrm{FiLM}}\,
\frac{1}{|\mathcal Z|}
\underset{(G,r)\ \mathrm{valid}}{\operatorname{mean}}
\left(\lVert\gamma_{\mathrm{raw}}(c_{G,r})\rVert_2^2
+\lVert\beta(c_{G,r})\rVert_2^2\right).
\label{eq:film-regularizer}
\end{equation}
Here $|\mathcal Z|$ is the element vocabulary size.

\subsection{Set objective and assignment}
\label{app:den-loss}

The clean-state prediction loss is
\begin{equation}
\begin{aligned}
\mathcal L_{\mathrm{sym}}
={}&\mathcal L_G+w_K\mathcal L_K+w_m\mathcal L_{\mathrm{mask}}
+w_r\mathcal L_r+w_Z\mathcal L_Z\\
&+s_{\mathrm{aux}}\left(
w_{\mathrm{cap}}\mathcal L_{\mathrm{cap}}
+w_N\mathcal L_N
+w_{\mathrm{comp}}\mathcal L_{\mathrm{comp}}
\right).
\end{aligned}
\label{eq:loss}
\end{equation}
The global losses are the clean-target cross-entropies
$\mathcal L_G=-\log p_\theta(G_0\mid x_t)$ and
$\mathcal L_K=-\log p_\theta(K_0\mid x_t)$ for the predicted space-group and
orbit-count distributions, respectively. For the slot losses, a Hungarian
assignment aligns the $M$ predictions with the $K_0$ clean occupations and
$M-K_0$ padding targets. Writing $\pi(j)$ for the predicted slot matched to
target $j$ and $\mathfrak S_M$ for the set of permutations of the $M$ slots,
the optimal assignment and its costs are
\begin{equation}
\begin{aligned}
&\pi^\star
=\arg\min_{\pi\in\mathfrak S_M}
\left[
\sum_{j=1}^{K_0}C_{\mathrm{occ}}\bigl(\pi(j),j\bigr)
+\sum_{j=K_0+1}^{M}C_{\mathrm{pad}}\bigl(\pi(j)\bigr)
\right],\\
&C_{\mathrm{occ}}(i,j)
=-w_m\log p_i^{\mathrm{act}}(1\mid x_t)
-w_r\log p_i(r_j\mid G_0,x_t)
-w_Z\log p_i(Z_j\mid G_0,r_j,x_t),\\
&C_{\mathrm{pad}}(i)
=-w_m\log p_i^{\mathrm{act}}(0\mid x_t).
\end{aligned}
\label{eq:symbolic-assignment}
\end{equation}
Here $C_{\mathrm{occ}}$ is the cost of assigning a predicted slot to a clean
occupation, whereas $C_{\mathrm{pad}}$ is the cost of assigning it to a padding
target.
After assignment, $\mathcal L_{\mathrm{mask}}$ is the binary cross-entropy
between each slot's active-state prediction and its matched target. The losses
$\mathcal L_r$ and $\mathcal L_Z$ are cross-entropies
between the predicted row and element distributions and their respective clean
targets over the matched occupied slots.

Let $\widehat m_i=p_i^{\mathrm{act}}(1\mid x_t)$ denote the predicted
probability that slot $i$ is active. The auxiliary losses are computed from the
expected row occupations
$\widehat n_r=\sum_i \widehat m_i p_i(r\mid G_0)$:
\begin{align}
&\begin{aligned}
&\mathcal L_{\mathrm{cap}}
=\sum_r\operatorname{ReLU}\!\left(\widehat n_r-u_{G_0,r}\right)^2,\\
&\mathcal L_N=(\widehat N-N_{\mathrm{sym},0})^2,\\
&\text{where}\quad
\widehat N=\sum_i \widehat m_i\sum_r
p_i(r\mid G_0)\mu_{G_0,r}.
\end{aligned}
\label{eq:auxcap}
\\
&\begin{aligned}
&\mathcal L_{\mathrm{comp}}
=\sum_a(\widehat C_a-C_{0,a})^2,\\
&\text{where}\quad
\widehat C_a=\sum_i \widehat m_i\sum_r
p_i(r\mid G_0)\mu_{G_0,r}p_i(a\mid G_0,r).
\end{aligned}
\label{eq:auxcomp}
\end{align}
Here $u_{G_0,r}$ is the permitted occupation count, $\mu_{G_0,r}$ is the
Wyckoff multiplicity, and $\widehat N$ and $N_{\mathrm{sym},0}$ are the expected
and clean conventional-cell atom counts, respectively. Likewise, $\widehat C_a$
and $C_{0,a}$ are the expected and clean multiplicity-weighted counts of element
$a$. The decoder in Appendix~\ref{app:decoder} separately enforces the
row-capacity limits and its fixed atom-count cap.

\subsection{Training settings}
\label{app:den-film-stage}

The symbolic denoiser is trained with Eq.~\eqref{eq:loss}. The row-conditioned
element head is then refined while all other parameters are frozen, using
\begin{equation}
\mathcal L_{\mathrm{refine}}
=\mathcal L_{\mathrm{sym}}+\mathcal L_{\mathrm{FiLM}}.
\label{eq:film-refinement}
\end{equation}
The reported configurations use dataset-specific capacity rules and symbolic
training schedules. Under the multiplicity-based rule, a valid row has capacity
\[
u_{G,r}^{\mathrm{mult}}=
\begin{cases}
1, & \operatorname{dof}_{G,r}=0,\\
\max\!\left\{1,\left\lfloor N_{\max}/\mu_{G,r}\right\rfloor\right\},
& \operatorname{dof}_{G,r}>0.
\end{cases}
\]
On MP-20, \ours{} uses training-set 99th-percentile row capacities, with
training budgets of 100,000 backbone steps followed by 20,000 element-head
refinement steps. \ours{}-II uses multiplicity-based capacities and budgets of
150,000 and 40,000 steps, respectively. On MPTS-52, both models use
multiplicity-based capacities: \ours{} has budgets of 150,000 backbone and
40,000 refinement steps, and \ours{}-II has budgets of 100,000 and 20,000 steps.
For each stage, we select the checkpoint with the lowest validation loss.

Table~\ref{tab:denoiser-hp} lists the common architecture and optimization
settings, with the dataset-specific orbit-slot counts.

\begin{table}[H]
\centering
\caption{Symbolic-denoiser architecture and optimization settings.}
\label{tab:denoiser-hp}
\scriptsize
\setlength{\tabcolsep}{3pt}
\begin{tabular}{@{}
>{\raggedright\arraybackslash}p{0.19\textwidth}
>{\raggedright\arraybackslash}p{0.16\textwidth}
>{\raggedright\arraybackslash}p{0.20\textwidth}
>{\raggedright\arraybackslash}p{0.36\textwidth}
@{}}
\toprule
\textbf{Architecture} & \textbf{Value}
& \textbf{Optimization} & \textbf{Value} \\
\midrule
hidden width & 512
& optimizer & AdamW \\
layers / attention heads & 8 / 8
& peak learning rate & $3\times10^{-4}$ \\
FFN width / dropout & $4d_h$ / 0
& weight decay / grad. clip & 0.01 / 1.0 \\
time / $K$ / mask / element emb. & 128 / 32 / 16 / 64
& batch size & 256 \\
orbit slots $M$ (MP-20 / MPTS-52) & 20 / 52
& learning-rate schedule & 2,000-step warmup; cosine to $0.05\times$ peak \\
FiLM hidden width & 128
& & \\
\midrule
core loss weights & $w_K=w_m=w_r=w_Z=1$
& auxiliary ramp & 5,000 steps \\
auxiliary weights
& $w_{\mathrm{cap}}=0.1,\ w_N=w_{\mathrm{comp}}=0.01$
& $\lambda_{\mathrm{FiLM}}$ & $10^{-4}$ \\
\bottomrule
\end{tabular}
\end{table}

\section{Legality-Constrained Decoding}
\label{app:decoder}

\subsection{Constrained target}

Fix a noisy state $x_t$, and abbreviate the denoiser outputs in
Eq.~\eqref{eq:heads} as $p_G(G)$, $p_K(K)$, active probabilities $\pi_i$,
row distributions $p_i(r\mid G)$, and element conditionals
$p_i(Z\mid G,r)$. For a clean symbolic state
$x_0=(G,K,A,\mathbf r,\mathbf Z)$ with active-slot set
$A\subseteq\{1,\ldots,M\}$, the factorized joint distribution is
\begin{equation}
p_\theta^{\mathrm{fac}}(x_0\mid x_t)
=p_G(G)\,p_K(K)
\prod_{i\in A}\pi_i\,p_i(r_i\mid G)\,p_i(Z_i\mid G,r_i)
\prod_{i\notin A}(1-\pi_i).
\label{eq:decoder-factorized}
\end{equation}
Its joint score is $S_\theta(x_0\mid x_t)=
\log p_\theta^{\mathrm{fac}}(x_0\mid x_t)$. Given the legal set
$\mathcal X_{\mathrm{legal}}$ from Section~\ref{sec:decoding}, the stochastic decoder
targets the constrained distribution
\begin{equation}
\widetilde p_\theta(x_0\mid x_t)
=
\frac{p_\theta^{\mathrm{fac}}(x_0\mid x_t)
\mathbf 1[x_0\in\mathcal X_{\mathrm{legal}}]}
{\sum_{x\in\mathcal X_{\mathrm{legal}}}p_\theta^{\mathrm{fac}}(x\mid x_t)}.
\label{eq:decoder-target}
\end{equation}
For the decoder comparison, we use a constrained maximum a posteriori (MAP)
comparator that retains the top five candidates under $p_G$ and $p_K$,
respectively, and performs deterministic constrained search over the resulting
$G$--$K$ pairs.

\subsection{Elementary-symmetric stochastic proposal}

Write $\omega_i=\pi_i/(1-\pi_i)=\exp(a_i)$, where $a_i$ is the active
logit. The elementary symmetric polynomial
\begin{equation}
e_k(\boldsymbol\omega)
=\sum_{\substack{A\subseteq\{1,\ldots,M\}\\|A|=k}}
\prod_{i\in A}\omega_i
\end{equation}
normalizes the distribution over size-$k$ active sets. It is evaluated by the
recursion
\begin{equation}
F_i(k)=F_{i+1}(k)+\omega_iF_{i+1}(k-1),
\qquad
F_{M+1}(0)=1,\quad F_{M+1}(k>0)=0,
\label{eq:decoder-esym}
\end{equation}
for which $F_1(k)=e_k(\boldsymbol\omega)$. The same table gives the conditional
inclusion probability
\begin{equation}
\Pr(i\in A\mid k\text{ active slots remain})
=\frac{\omega_iF_{i+1}(k-1)}{F_i(k)}.
\label{eq:decoder-subset}
\end{equation}
The proposal draws $G\sim p_G$, draws
$K$ with probability proportional to $p_K(K)e_K(\boldsymbol\omega)$,
draws $A$ from
\begin{equation}
q(A\mid K)=
\frac{\prod_{i\in A}\omega_i}{e_K(\boldsymbol\omega)}
\quad\text{for }|A|=K,
\end{equation}
and then draws each $r_i\sim p_i(\cdot\mid G)$.

\subsection{Exactness}

We prove Proposition~\ref{prop:exact}. Let $s=(G,K,A,\mathbf r)$ denote the symbolic
skeleton, and let $\mathcal S_{\mathrm{legal}}$ be the projection of
$\mathcal X_{\mathrm{legal}}$ onto these variables. Because
$\sum_{Z\in\mathcal Z}p_i(Z\mid G,r_i)=1$ for every occupied slot, marginalizing
Eq.~\eqref{eq:decoder-factorized} over $\mathbf Z$ gives
\begin{equation}
p_\theta^{\mathrm{skel}}(s\mid x_t)
=c_0\,p_G(G)\,p_K(K)
\prod_{i\in A}\omega_i\,p_i(r_i\mid G),
\qquad
c_0=\prod_{i=1}^{M}(1-\pi_i).
\label{eq:decoder-skeleton}
\end{equation}
Let
$Z_K=\sum_k p_K(k)e_k(\boldsymbol\omega)$. One proposal attempt has density
\begin{align}
q(s)
&=p_G(G)
\frac{p_K(K)e_K(\boldsymbol\omega)}{Z_K}
\frac{\prod_{i\in A}\omega_i}{e_K(\boldsymbol\omega)}
\prod_{i\in A}p_i(r_i\mid G) \notag\\
&=\frac{p_\theta^{\mathrm{skel}}(s\mid x_t)}{Z_Kc_0}.
\label{eq:decoder-proposal-density}
\end{align}
The elementary-symmetric factors therefore cancel, and the proposal is
proportional to the unconstrained skeleton marginal. Conditioning an attempt on
$s\in\mathcal S_{\mathrm{legal}}$ cancels the remaining constant:
\begin{equation}
\Pr(s\mid s\in\mathcal S_{\mathrm{legal}})
=
\frac{p_\theta^{\mathrm{skel}}(s\mid x_t)\mathbf 1[s\in\mathcal S_{\mathrm{legal}}]}
{\sum_{s'\in\mathcal S_{\mathrm{legal}}}p_\theta^{\mathrm{skel}}(s'\mid x_t)}.
\end{equation}
After acceptance, drawing each element from
$p_i(\cdot\mid G,r_i)$ restores the full joint distribution in
Eq.~\eqref{eq:decoder-target}. Thus the accepted output is an exact sample from the
constrained target without evaluating its global normalizer.

\subsection{Sampling procedure}

For each reverse step, the elementary-symmetric table is computed once. For at most
$R$ attempts, the decoder draws $(G,K,A,\mathbf r)$ from the proposal above and
checks whether the resulting skeleton lies in $\mathcal S_{\mathrm{legal}}$. On
acceptance, it draws $\mathbf Z$ from the element conditionals and returns the
complete clean symbolic state; otherwise it redraws the entire skeleton. If none
of the $R$ attempts is accepted, the decoder reports a rejection to the reverse
sampler.

Algorithm~\ref{alg:full-generation} summarizes the generation path with
accepted symbolic proposals.

\begin{algorithm}[htb]
\caption{\ours{} generation}
\label{alg:full-generation}
\begin{algorithmic}[1]
\Require Symbolic denoiser, forward kernels $q_t$, sampled levels $t_1>\cdots>t_J$, geometry model
\State Draw $x_{t_1}$ from the terminal prior.
\For{$j=1,\ldots,J$}
  \State Predict the clean-state distributions from $x_{t_j}$.
  \State Decode a clean symbolic state $\hat x_0^{(t_j)}$ using the procedure above.
  \If{$j<J$}
    \State Draw $x_{t_{j+1}}\sim q_{t_{j+1}}(\cdot\mid\hat x_0^{(t_j)})$.
  \EndIf
\EndFor
\State Remove inactive slots and slot order from $\hat x_0^{(t_J)}$ to obtain $P$.
\State Generate the lattice and orbit anchors conditioned on $P$ (Appendix~\ref{app:realization}).
\State Expand the anchors through the Wyckoff operations of $P$ and return the crystal.
\end{algorithmic}
\end{algorithm}

\section{Symmetry-Constrained Geometry Generation}
\label{app:realization}

The geometry model receives an accepted protostructure
$P=(G,K,\{(Z_i,r_i)\}_{i=1}^{K})$ from the symbolic stage and generates a
lattice together with the free fractional coordinates of its occupied Wyckoff
orbits. Its symmetry-constrained diffusion of the lattice representation and orbit-anchor
coordinates builds on DiffCSP++ \citep{diffcsppp}. Here $G$ determines the
lattice subspace, each $r_i$ determines the orbit chart and expansion
operators, and the denoiser receives the corresponding element and
shared-codebook embeddings.

\subsection{Geometry denoiser}

The denoiser is an atom-level message-passing network over the expanded unit
cell. Periodicity enters through Fourier features of fractional-coordinate
differences and the current lattice representation. The network also receives the
diffusion time and protostructure conditioning. It produces a coordinate
denoising field for every atom and a pooled lattice-noise prediction.

\subsection{Lattice diffusion under crystal-family constraints}

Let $b\in\mathbb R^6$ be the coordinates of
$\log\sqrt{LL^{\top}}$ in an orthonormal basis for symmetric matrices. This
rotation-invariant representation is projected onto the crystal-family subspace associated
with $G$ by the affine map $\Pi_G(b)=\Pi_G^{\mathrm{lin}}b+d_G$.
The forward process is
\begin{equation}
b_t
=\Pi_G\!\left(
\sqrt{\bar\alpha_t}\,b_0
+\sqrt{1-\bar\alpha_t}\,\epsilon
\right),
\qquad
\epsilon_G=\Pi_G^{\mathrm{lin}}\epsilon,
\quad
\epsilon\sim\mathcal N(0,I),
\label{eq:lattice-forward}
\end{equation}
and the lattice head minimizes
\begin{equation}
\mathcal L_{\mathrm{lat}}
=\mathbb E\left[
\left\lVert
\epsilon_G-\widehat\epsilon_\theta(b_t,X_t,P,t)
\right\rVert_2^2
\right].
\label{eq:lattice-loss}
\end{equation}
Projection is applied after every noising and reverse update. With the usual
DDPM notation \citep{ddpm} and $z_t\sim\mathcal N(0,I_6)$, the reverse step is
\begin{equation}
b_{t-1}
=\Pi_G\!\left[
\frac{1}{\sqrt{\alpha_t}}
\left(
b_t-\frac{1-\alpha_t}{\sqrt{1-\bar\alpha_t}}\,
\widehat\epsilon_\theta
\right)
+\sigma_{t,\mathrm{lat}}\,z_t
\right],
\qquad
\sigma_{t,\mathrm{lat}}^2
=\beta_t\frac{1-\bar\alpha_{t-1}}{1-\bar\alpha_t}.
\label{eq:lattice-reverse}
\end{equation}

\subsection{Anchor-coordinate score matching}

Each occupied Wyckoff orbit is parameterized by one anchor coordinate. Its
Wyckoff affine operators deterministically expand that anchor into the full
orbit. During training, coordinate noise is mapped to the free anchor
coordinates, where the denoising target is evaluated. The predicted field is
mapped to the same coordinates for the loss. During sampling, the predicted
field is mapped in the same way before the updated anchors are re-expanded.

Let $u_{\mathrm{wn}}^*(X_t,X_0;\sigma_t)$ denote the wrapped-normal
denoising target in the orbit-anchor coordinates under the adopted noise
normalization, and let $\Pi_{\mathcal A(P)}$ map each predicted atomic field to
the corresponding anchor coordinates defined by $P$. The coordinate objective is
\begin{equation}
\mathcal L_{\mathrm{coord}}
=\mathbb E\left[
\left\lVert
\Pi_{\mathcal A(P)}\widehat u_\theta(X_t,b_t,P,t)
-u_{\mathrm{wn}}^*(X_t,X_0;\sigma_t)
\right\rVert_2^2
\right].
\label{eq:coordinate-loss}
\end{equation}
The reverse process uses a coordinate corrector followed by a joint predictor
step for the lattice and coordinates. After each update, orbit expansion
reconstructs the complete unit cell, so every intermediate structure obeys
the generated space group.

\subsection{Architecture and training settings}

A separate geometry model is trained on the training split of each dataset,
using conventional-cell lattices and orbit anchors. The symbolic configurations
within a dataset share its geometry-model weights. Both datasets use the common
geometry architecture and training settings in Table~\ref{tab:geometry-hp};
the smaller variant, \ourssmall{}, is evaluated on MP-20 in the efficiency
comparison.

\begin{table}[H]
\centering
\caption{Geometry model architecture and training settings.}
\label{tab:geometry-hp}
\footnotesize
\setlength{\tabcolsep}{4pt}
\begin{tabular}{@{}ll@{\qquad}ll@{}}
\toprule
\textbf{Architecture} & \textbf{Value}
& \textbf{Optimization / diffusion} & \textbf{Value} \\
\midrule
\ours{} (ours) width / layers & 512 / 6
& optimizer & AdamW \\
\ourssmall{} (ours) width / layers & 256 / 4
& learning rate / weight decay & $10^{-3}$ / $10^{-2}$ \\
& & batch size / grad. clip & 128 / 1.0 \\
& & warmup / training-step budget & 500 / 60,000 \\
& & lattice schedule & cosine, 1,000 levels \\
& & coordinate noise range & $[0.005,0.5]$, 1,000 levels \\
& & corrector coefficient & $10^{-5}$ \\
\bottomrule
\end{tabular}
\end{table}

\section{Evaluation Protocol}
\label{app:protocol}

\subsection{Datasets, relaxation, and references}
\label{app:protocol-chain}

We evaluate MP-20 \citep{cdvae} and MPTS-52 \citep{sgequidiff} separately, using
each dataset's training partition for novelty and
its test partition for coverage. Discovery rates in the main comparisons use a candidate budget of
approximately 10,000 per evaluated set. The MP-20 geometry-step comparison uses the 10,000 generation attempts per seed as its denominator (Appendix~\ref{app:geometry-steps}).

CHGNet \citep{chgnet} relaxes the generated structures. Structures from both datasets are also independently relaxed with MACE-MP-0
\citep{mace}; the two engines define separate evaluation scales.
Energies above the hull are evaluated against the fixed, MP2020-corrected
Materials Project \citep{materialsproject} DFT snapshot dated 2023-02-07.
The stability thresholds are $\tau=0$ and $0.1$~eV/atom.

\subsection{Discovery and symmetry}
\label{app:provenance}

Structural uniqueness is evaluated within each generated set and novelty
against the corresponding training partition, using pymatgen's
\texttt{StructureMatcher} \citep{pymatgen} with its default matching tolerances.
SUN@$\tau$ counts candidates that are unique, novel, and satisfy
$E_{\mathrm{hull}}\leq\tau$, divided by the candidate budget.
SSUN@$\tau$ further requires a successfully detected non-$P1$ space group
after relaxation. Space groups are detected with spglib \citep{spglib} via
\texttt{SpacegroupAnalyzer}, using \texttt{symprec}$=0.1$~\AA{} and an angular
tolerance of $5^{\circ}$. Failed symmetry identifications do not enter the SSUN
numerator. Stable@$\tau$ in the ablation tables counts all stable candidates
over the same candidate budget, without the uniqueness and novelty requirements.

The space-group distribution in Section~\ref{sec:sg-distribution} instead uses
as-generated coordinates and the full MP-20 dataset as the reference, with the
same symmetry tolerances. The $P1$ fractions in Tables~\ref{tab:mp20-quality} and~\ref{tab:mpts52-quality}
use CHGNet-relaxed structures after conventional-cell standardization:
their numerators count $P1$ structures and their denominators count
successful symmetry identifications.
These rates are computed within each run before averaging.

\subsection{Validity, distributional fidelity, and geometric quality}
\label{app:quality-reference}

Tables~\ref{tab:mp20-quality} and~\ref{tab:mpts52-quality} describe CHGNet-relaxed structures. Structural
validity tests the minimum interatomic separation; compositional validity uses
the CDVAE-compatible oxidation-state and electronegativity checks, including
the alloy convention. Distributional distances use structurally and
compositionally valid relaxed structures, with the MP-20 test partition and
the MPTS-52 training partition as references. We use Wasserstein distances for
mass density and the number of distinct elements, and Jensen--Shannon divergence
in bits for the pooled distribution of Wyckoff degrees of freedom, counting
each occupied orbit once. Wyckoff positions are identified at
\texttt{symprec}$=0.01$~\AA{} and an angular tolerance of $5^{\circ}$.

Train--test distribution distances are reported in
Table~\ref{tab:reference-split-distances}. Across all three statistics, the
train--test differences are negligible for MP-20 but pronounced for MPTS-52,
consistent with the random split of MP-20 and the chronological split of MPTS-52.

\begin{table}[htb]
\centering
\caption{Distribution distances between the training and test splits,
computed using all available reference features.}
\label{tab:reference-split-distances}
\begingroup
\small
\setlength{\tabcolsep}{8pt}
\renewcommand{\arraystretch}{1.08}
\begin{tabular}{@{}lccc@{}}
\toprule
Dataset & $d_\rho$ (g/cm$^3$) & $d_{N_{\mathrm{el}}}$ & $d_{\mathrm{Wyckoff}}$ (bits) \\
\midrule
MP-20 & 0.048616 & 0.011125 & 0.000130 \\
MPTS-52 & 0.892426 & 0.633220 & 0.026667 \\
\bottomrule
\end{tabular}
\endgroup

\end{table}

Coverage recall and
precision use joint structural and compositional fingerprint matching against
the test partition \citep{cdvae}. On MP-20, validity and coverage retain their evaluated-set
denominators. On MPTS-52, structural validity, compositional validity, and
coverage precision are normalized by the 10,000-candidate budget. Coverage
recall always uses the test reference. Median $E_{\mathrm{hull}}$ summarizes
the relaxed structures with evaluable hull energies.

For RMSD, we pair each as-generated structure with its CHGNet-relaxed endpoint
using the same MatterGen-derived matching procedure \citep{mattergen} on both
datasets. The generated structure is refined at
\texttt{symprec}$=0.01$~\AA{} and $5^{\circ}$ angular tolerance, converted to a
primitive cell, and LLL-reduced; the relaxed endpoint is used as stored.
Primitive-cell matching and supercell attempts are enabled, with scaling
disabled and subsets disallowed. We report the mean
root-mean-square displacement (RMSD, in \AA{}) over matched structure pairs,
and Matched is the percentage of all evaluated pairs for which correspondence is found.

\subsection{First-principles validation on MP-20}
\label{app:vasp}

For each method, we form a unique-and-novel pool using its CHGNet-relaxed
structures and select the first 2,000 candidates in a fixed random ordering of
that pool. Multi-run methods use generation seed 0; for the other baselines,
we use their publicly available generated samples. VASP (version 6.4.2) starts from the corresponding original
generated structures, using the Materials Project relaxation settings supplied
by the standard \texttt{MPRelaxSet} input generator. Energies receive the MP2020 compatibility corrections
before evaluation against the same reference hull. The validation window is selected without ranking
candidates by their CHGNet energies.

Let $N_{\mathrm{gen}}$ be the original candidate population,
$N_{\mathrm{UN}}$ the size of its CHGNet unique-and-novel pool, and
$n_{\mathrm{stable}}$ the number of converged compounds in the VASP window
with $E_{\mathrm{hull}}\leq0.1$~eV/atom. Compounds must contain at least two
distinct elements. The reported rate is
\[
\mathrm{SUN\mbox{-}VASP}
=\frac{N_{\mathrm{UN}}}{N_{\mathrm{gen}}}
\frac{n_{\mathrm{stable}}}{2,000}.
\]
SSUN-VASP uses the same expression with the numerator additionally restricted
to non-$P1$ VASP endpoints detected at \texttt{symprec}$=0.1$~\AA{}.
The unique-and-novel condition is the CHGNet pool selection; VASP supplies the
stability and final-symmetry tests.

\subsection{Aggregation and timing}
\label{app:protocol-timing}

Each reported \ours{} configuration and SGEquiDiff is evaluated over five
generation seeds; for each of the other main-table baselines, we evaluate
their publicly available sample set. We average the per-run statistics without pooling structures, including
the per-run medians of $E_{\mathrm{hull}}$. Appendix~\ref{app:variability}
reports the corresponding means and sample standard deviations. The VASP window
is a single evaluation per row and is not assigned a sampling-seed standard
deviation.

Timing covers complete generation through the as-generated structure,
including both symbolic and geometric generation for \ours{}.
Generation rate is the number of complete structures
per second. All measurements are performed on the same machine equipped with
an NVIDIA H100 GPU (80 GB) and 32 CPU processes.

\subsection{Geometry sampling steps}
\label{app:geometry-steps}

\begin{figure}[htb]
\centering
\includegraphics[width=4.25in]{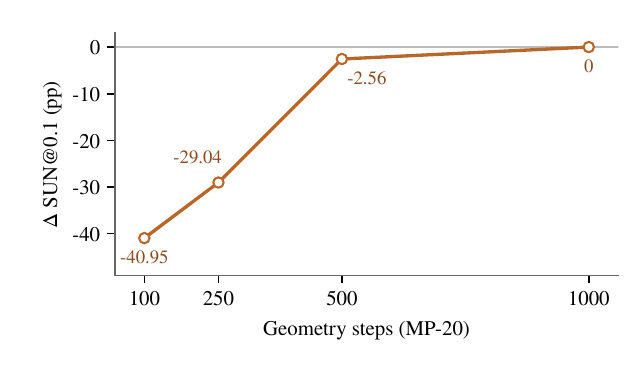}
\caption{Geometry sampling steps on MP-20. Changes in SUN@0.1 relative to 1,000 geometry steps are averaged over five paired generation seeds of \ours{}. Rates use the 10,000-candidate budget as the denominator; the symbolic samples and geometry model are held fixed.}
\label{fig:geometry-steps}
\end{figure}

We compare 1,000, 500, 250, and 100 geometry sampling steps for \ours{} on
MP-20, using the same symbolic samples and geometry model across schedules.
Symbolic sampling retains 200 noise levels. Results are means over five
paired generation seeds. All schedules use a fixed budget of 10,000
candidates per seed, including unsuccessful structure construction.
The mean SUN@0.1 rates at 1,000 and 500 steps are 41.49\% and 38.93\%,
respectively, so halving the geometry sampling steps retains approximately
93.8\% of the discovery rate. Figure~\ref{fig:geometry-steps} shows the
complete comparison.

\newpage
\section{Independent Evaluation with MACE-MP-0}
\label{app:mace}

We independently relax the MP-20 and MPTS-52 generated sets with MACE-MP-0.
For each dataset, we repeat the discovery analysis using the reference data and definitions in
Appendix~\ref{app:protocol}. The symmetry test uses
\texttt{symprec}$=0.1$~\AA{} and $5^{\circ}$ angular tolerance on the
MACE-MP-0 endpoints. Both \ours{} configurations lead the compared
symmetry-aware baselines across the reported discovery metrics.

\begin{table}[htbp]
\centering
\caption{MP-20 discovery under MACE-MP-0 (\%; higher is better).
Both \ours{} configurations and SGEquiDiff show mean $\mathbin{\pm}$ sample standard deviation
over five generation seeds; for the other methods, we evaluate publicly available
generated samples. Thresholds @0 and @0.1 are in eV/atom. All values in this
table use the MACE-MP-0 evaluation scale.}
\label{tab:mp20-mace-discovery}
\begingroup
\small
\setlength{\tabcolsep}{4pt}
\renewcommand{\arraystretch}{1.08}
\begin{tabular*}{\linewidth}{@{\extracolsep{\fill}}lrrrr@{}}
\toprule
Method & SUN@0 & SSUN@0 & SUN@0.1 & SSUN@0.1 \\
\midrule
DiffCSP++ & $7.93$ & $7.87$ & $34.69$ & $34.40$ \\
SymmCD & $7.88$ & $7.79$ & $30.04$ & $29.69$ \\
WyFormer & $6.43$ & $6.40$ & $29.66$ & $29.09$ \\
SGEquiDiff & $9.11 \pm 0.23$ & $8.98 \pm 0.24$ & $35.28 \pm 0.27$ & $34.51 \pm 0.28$ \\
\midrule
\ours{} (ours) & $9.55 \pm 0.36$ & $9.44 \pm 0.37$ & $38.17 \pm 0.32$ & $37.24 \pm 0.33$ \\
\ours{}-II (ours) & $9.61 \pm 0.45$ & $9.53 \pm 0.45$ & $38.69 \pm 0.53$ & $37.67 \pm 0.62$ \\
\bottomrule
\end{tabular*}
\endgroup

\end{table}

Tables~\ref{tab:mpts52-mace-discovery} and~\ref{tab:mpts52-mace-quality}
report discovery and quality results on MPTS-52 under MACE-MP-0.
Quality metrics follow the definitions and reference sets in
Appendix~\ref{app:protocol}, using MACE-MP-0-relaxed endpoints.

\begin{table}[htb]
\centering
\caption{MPTS-52 discovery under MACE-MP-0 (\%; higher is better).
Reporting conventions follow Table~\ref{tab:mp20-mace-discovery}.}
\label{tab:mpts52-mace-discovery}
\begingroup
\small
\setlength{\tabcolsep}{4pt}
\renewcommand{\arraystretch}{1.08}
\begin{tabular*}{\linewidth}{@{\extracolsep{\fill}}lrrrr@{}}
\toprule
Method & SUN@0 & SSUN@0 & SUN@0.1 & SSUN@0.1 \\
\midrule
SymmCD & $6.21$ & $6.07$ & $19.61$ & $19.36$ \\
WyFormer & $2.98$ & $2.84$ & $9.22$ & $8.95$ \\
SGEquiDiff & $5.46 \pm 0.19$ & $5.32 \pm 0.19$ & $22.33 \pm 0.27$ & $21.83 \pm 0.24$ \\
\midrule
\ours{} (ours) & $7.48 \pm 0.24$ & $7.40 \pm 0.25$ & $30.22 \pm 0.49$ & $29.87 \pm 0.47$ \\
\ours{}-II (ours) & $6.79 \pm 0.18$ & $6.73 \pm 0.18$ & $28.52 \pm 0.42$ & $28.20 \pm 0.45$ \\
\bottomrule
\end{tabular*}
\endgroup

\end{table}

\begin{table}[htb]
\centering
\caption{MPTS-52 validity and distributional fidelity under MACE-MP-0.
Distributional distances use valid relaxed structures and the training reference.
Reporting conventions follow Table~\ref{tab:mpts52-mace-discovery}.}
\label{tab:mpts52-mace-quality}
\begingroup
\small
\setlength{\tabcolsep}{2pt}
\renewcommand{\arraystretch}{1.08}
\begin{tabular*}{\linewidth}{@{\extracolsep{\fill}}lrrrrr@{}}
\toprule
Metric & SymmCD & WyFormer & SGEquiDiff & \ours{} & \ours{}-II \\
\midrule
Struct. val. (\%) $\uparrow$ & $87.150$ & $84.360$ & $97.764 \pm 0.218$ & $99.998 \pm 0.004$ & $99.998 \pm 0.004$ \\
Comp. val. (\%) $\uparrow$ & $69.94$ & $64.17$ & $78.32 \pm 0.21$ & $75.03 \pm 0.33$ & $76.93 \pm 0.23$ \\
$d_\rho$ (g/cm$^3$) $\downarrow$ & $0.1715$ & $0.7190$ & $0.2598 \pm 0.0193$ & $0.2809 \pm 0.0133$ & $0.1899 \pm 0.0127$ \\
$d_{N_{\mathrm{el}}}$ $\downarrow$ & $0.5629$ & $0.2398$ & $0.1482 \pm 0.0038$ & $0.0504 \pm 0.0038$ & $0.0201 \pm 0.0081$ \\
$d_{\mathrm{Wyckoff}}$ (bits) $\downarrow$ & $0.0155$ & $0.0698$ & $0.0545 \pm 0.0015$ & $0.0252 \pm 0.0023$ & $0.0168 \pm 0.0012$ \\
Cov. R (\%) $\uparrow$ & $99.80$ & $99.89$ & $99.96 \pm 0.02$ & $99.85 \pm 0.05$ & $99.85 \pm 0.04$ \\
Cov. P (\%) $\uparrow$ & $85.20$ & $81.20$ & $94.06 \pm 0.27$ & $98.56 \pm 0.16$ & $98.32 \pm 0.10$ \\
$P1$ (\%) $\downarrow$ & $1.89$ & $4.91$ & $3.79 \pm 0.17$ & $1.92 \pm 0.08$ & $2.06 \pm 0.14$ \\
\shortstack[l]{Median $E_{\mathrm{hull}}$\\(eV/atom) $\downarrow$} & $0.225$ & $0.395$ & $0.185 \pm 0.002$ & $0.108 \pm 0.002$ & $0.114 \pm 0.003$ \\
\bottomrule
\end{tabular*}
\endgroup

\end{table}

\clearpage
\section{Across-Run Variability}
\label{app:additional-evaluation}
\label{app:variability}

For CHGNet evaluation, we report means and sample standard deviations across five generation seeds.
Results for the remaining methods and VASP evaluations are given in the main-text tables.

\begin{table}[htb]
\centering
\caption{Sampling variability on MP-20 and MPTS-52
(mean $\pm$ sample standard deviation). Metrics and units follow
Tables~\ref{tab:mp20-discovery}--\ref{tab:rmsd}.}
\label{tab:variability}
\label{tab:mean-sd-summary}
\begingroup
\small
\setlength{\tabcolsep}{2pt}
\renewcommand{\arraystretch}{1.08}
\begin{tabular*}{\linewidth}{@{\extracolsep{\fill}}lrrrrr@{}}
\toprule
\multicolumn{6}{@{}l}{\textbf{(a) Material discovery}} \\
Method & SUN@0 (\%) & SSUN@0 (\%) & SUN@0.1 (\%) & \multicolumn{2}{r}{SSUN@0.1 (\%)} \\
\midrule
\multicolumn{6}{@{}l}{\emph{MP-20}} \\
SGEquiDiff & $8.45 \pm 0.34$ & $8.31 \pm 0.35$ & $38.00 \pm 0.32$ & \multicolumn{2}{r}{$36.38 \pm 0.36$} \\
\midrule
\ours{} (ours) & $9.39 \pm 0.14$ & $9.31 \pm 0.12$ & $41.49 \pm 0.34$ & \multicolumn{2}{r}{$40.12 \pm 0.25$} \\
\ours{}-II (ours) & $9.18 \pm 0.30$ & $9.10 \pm 0.29$ & $41.32 \pm 0.51$ & \multicolumn{2}{r}{$39.87 \pm 0.52$} \\
\addlinespace
\multicolumn{6}{@{}l}{\emph{MPTS-52}} \\
SGEquiDiff & $5.17 \pm 0.20$ & $4.94 \pm 0.18$ & $26.29 \pm 0.36$ & \multicolumn{2}{r}{$24.33 \pm 0.35$} \\
\midrule
\ours{} (ours) & $7.30 \pm 0.42$ & $7.16 \pm 0.44$ & $34.20 \pm 0.53$ & \multicolumn{2}{r}{$33.05 \pm 0.48$} \\
\ours{}-II (ours) & $6.65 \pm 0.19$ & $6.58 \pm 0.19$ & $31.85 \pm 0.34$ & \multicolumn{2}{r}{$30.85 \pm 0.39$} \\
\midrule
\addlinespace
\multicolumn{6}{@{}l}{\textbf{(b) Validity and distributional fidelity}} \\
Method & Struct. (\%) & Comp. (\%) & $d_{\rho}$ & $d_{N_{\mathrm{el}}}$ & $d_{\mathrm{Wyckoff}}$ \\
\midrule
\multicolumn{6}{@{}l}{\emph{MP-20}} \\
SGEquiDiff & $99.50 \pm 0.10$ & $86.13 \pm 0.43$ & $0.1761 \pm 0.0160$ & $0.1195 \pm 0.0102$ & $0.0781 \pm 0.0029$ \\
\midrule
\ours{} (ours) & $99.97 \pm 0.01$ & $83.36 \pm 0.19$ & $0.1025 \pm 0.0245$ & $0.0518 \pm 0.0085$ & $0.0340 \pm 0.0032$ \\
\ours{}-II (ours) & $99.95 \pm 0.02$ & $83.30 \pm 0.54$ & $0.1225 \pm 0.0358$ & $0.0627 \pm 0.0052$ & $0.0344 \pm 0.0012$ \\
\addlinespace
\multicolumn{6}{@{}l}{\emph{MPTS-52}} \\
SGEquiDiff & $98.36 \pm 0.16$ & $79.91 \pm 0.17$ & $0.2400 \pm 0.0228$ & $0.1479 \pm 0.0031$ & $0.1266 \pm 0.0023$ \\
\midrule
\ours{} (ours) & $99.91 \pm 0.01$ & $75.02 \pm 0.32$ & $0.2618 \pm 0.0149$ & $0.0501 \pm 0.0037$ & $0.0626 \pm 0.0024$ \\
\ours{}-II (ours) & $99.90 \pm 0.06$ & $76.90 \pm 0.24$ & $0.1779 \pm 0.0130$ & $0.0202 \pm 0.0083$ & $0.0501 \pm 0.0033$ \\
\midrule
\addlinespace
\multicolumn{6}{@{}l}{\textbf{Validity and distributional fidelity (continued)}} \\
Method & Cov. R (\%) & Cov. P (\%) & P1 (\%) & \multicolumn{2}{r}{\shortstack{Median $E_{\mathrm{hull}}$\\(eV/atom)}} \\
\midrule
\multicolumn{6}{@{}l}{\emph{MP-20}} \\
SGEquiDiff & $99.76 \pm 0.04$ & $99.14 \pm 0.06$ & $7.95 \pm 0.12$ & \multicolumn{2}{r}{$0.082 \pm 0.001$} \\
\midrule
\ours{} (ours) & $99.71 \pm 0.05$ & $99.31 \pm 0.10$ & $5.54 \pm 0.29$ & \multicolumn{2}{r}{$0.072 \pm 0.001$} \\
\ours{}-II (ours) & $99.70 \pm 0.07$ & $99.42 \pm 0.07$ & $5.42 \pm 0.12$ & \multicolumn{2}{r}{$0.060 \pm 0.003$} \\
\addlinespace
\multicolumn{6}{@{}l}{\emph{MPTS-52}} \\
SGEquiDiff & $99.94 \pm 0.02$ & $96.24 \pm 0.27$ & $10.09 \pm 0.27$ & \multicolumn{2}{r}{$0.161 \pm 0.001$} \\
\midrule
\ours{} (ours) & $99.83 \pm 0.07$ & $98.54 \pm 0.13$ & $4.59 \pm 0.13$ & \multicolumn{2}{r}{$0.093 \pm 0.003$} \\
\ours{}-II (ours) & $99.84 \pm 0.05$ & $98.35 \pm 0.06$ & $4.66 \pm 0.20$ & \multicolumn{2}{r}{$0.101 \pm 0.002$} \\
\midrule
\addlinespace
\multicolumn{6}{@{}l}{\textbf{(c) Geometric quality}} \\
 & \multicolumn{2}{c}{MP-20} & \multicolumn{3}{c}{MPTS-52} \\
\cmidrule(lr){2-3}\cmidrule(lr){4-6}
Method & RMSD (\AA) & Matched (\%) & RMSD (\AA) & \multicolumn{2}{r}{Matched (\%)} \\
\midrule
SGEquiDiff & $0.190 \pm 0.002$ & $91.38 \pm 0.10$ & $0.294 \pm 0.003$ & \multicolumn{2}{r}{$88.05 \pm 0.29$} \\
\midrule
\ours{} (ours) & $0.172 \pm 0.003$ & $96.18 \pm 0.21$ & $0.255 \pm 0.003$ & \multicolumn{2}{r}{$97.11 \pm 0.10$} \\
\ours{}-II (ours) & $0.164 \pm 0.003$ & $96.65 \pm 0.25$ & $0.239 \pm 0.003$ & \multicolumn{2}{r}{$96.95 \pm 0.22$} \\
\bottomrule
\end{tabular*}
\endgroup

\end{table}

\clearpage
\subsection{Detailed ablation results}
\label{app:full-ablation}

\begin{table}[htbp]
\centering
\caption{Detailed ablation results on MPTS-52 under CHGNet,
reported as mean $\pm$ sample standard deviation over five generation seeds.
Fixed $G$, all-to-all, and w/o FiLM are compared with \ours{};
MAP is compared with w/o FiLM using the same checkpoint.
Discovery and stability rates use the candidate budget; distributional distances
use valid relaxed structures and the training reference.}
\label{tab:full-ablation}
\begingroup
\maintablefont
\setlength{\tabcolsep}{2pt}
\renewcommand{\arraystretch}{1.06}
\begin{tabular*}{\linewidth}{@{\extracolsep{\fill}}lrrrrr@{}}
\toprule
Metric & \ours{} & Fixed $G$ & All-to-all & w/o FiLM & MAP \\
\midrule
\multicolumn{6}{@{}l}{\emph{Discovery and stability}} \\
SUN@0 (\%) $\uparrow$ & $7.30 \pm 0.42$ & $6.32 \pm 0.25$ & $6.71 \pm 0.38$ & $6.51 \pm 0.21$ & $5.35 \pm 0.19$ \\
SSUN@0 (\%) $\uparrow$ & $7.16 \pm 0.44$ & $6.21 \pm 0.28$ & $6.64 \pm 0.38$ & $6.41 \pm 0.23$ & $5.29 \pm 0.18$ \\
SUN@0.1 (\%) $\uparrow$ & $34.20 \pm 0.53$ & $29.43 \pm 0.38$ & $31.61 \pm 0.51$ & $32.91 \pm 0.48$ & $27.65 \pm 0.65$ \\
SSUN@0.1 (\%) $\uparrow$ & $33.05 \pm 0.48$ & $28.32 \pm 0.40$ & $30.90 \pm 0.50$ & $31.73 \pm 0.41$ & $27.06 \pm 0.64$ \\
Stable@0.1 (\%) $\uparrow$ & $51.87 \pm 0.73$ & $46.00 \pm 0.25$ & $50.68 \pm 0.56$ & $49.29 \pm 0.78$ & $46.73 \pm 0.42$ \\
\shortstack[l]{Median $E_{\mathrm{hull}}$\\(eV/atom) $\downarrow$} & $0.093 \pm 0.003$ & $0.118 \pm 0.001$ & $0.097 \pm 0.002$ & $0.102 \pm 0.003$ & $0.114 \pm 0.002$ \\
\midrule
\multicolumn{6}{@{}l}{\emph{Validity and coverage}} \\
Struct. val. (\%) $\uparrow$ & $99.91 \pm 0.01$ & $99.88 \pm 0.03$ & $99.94 \pm 0.02$ & $99.90 \pm 0.03$ & $99.96 \pm 0.03$ \\
Comp. val. (\%) $\uparrow$ & $75.02 \pm 0.32$ & $75.81 \pm 0.43$ & $74.88 \pm 0.26$ & $76.48 \pm 0.20$ & $79.74 \pm 0.36$ \\
Cov. R (\%) $\uparrow$ & $99.83 \pm 0.07$ & $99.91 \pm 0.02$ & $99.85 \pm 0.05$ & $99.81 \pm 0.05$ & $99.91 \pm 0.02$ \\
Cov. P (\%) $\uparrow$ & $98.54 \pm 0.13$ & $97.30 \pm 0.25$ & $98.44 \pm 0.06$ & $98.49 \pm 0.12$ & $97.51 \pm 0.17$ \\
\midrule
\multicolumn{6}{@{}l}{\emph{Distributional and geometric quality}} \\
$d_\rho$ (g/cm$^3$) $\downarrow$ & $0.2618 \pm 0.0149$ & $0.1631 \pm 0.0243$ & $0.4148 \pm 0.0474$ & $0.4053 \pm 0.0255$ & $0.3927 \pm 0.0493$ \\
$d_{N_{\mathrm{el}}}$ $\downarrow$ & $0.0501 \pm 0.0037$ & $0.1120 \pm 0.0077$ & $0.0685 \pm 0.0041$ & $0.0367 \pm 0.0065$ & $0.1386 \pm 0.0106$ \\
$d_{\mathrm{Wyckoff}}$ (bits) $\downarrow$ & $0.0626 \pm 0.0024$ & $0.0495 \pm 0.0026$ & $0.0573 \pm 0.0023$ & $0.0718 \pm 0.0021$ & $0.0500 \pm 0.0035$ \\
RMSD (\AA) $\downarrow$ & $0.255 \pm 0.003$ & $0.226 \pm 0.003$ & $0.247 \pm 0.001$ & $0.262 \pm 0.001$ & $0.233 \pm 0.003$ \\
\bottomrule
\end{tabular*}
\endgroup

\end{table}

\clearpage
\section{Generated crystal examples}
\label{app:generated-examples}

\begin{figure}[htbp]
\centering
\includegraphics[width=\linewidth]{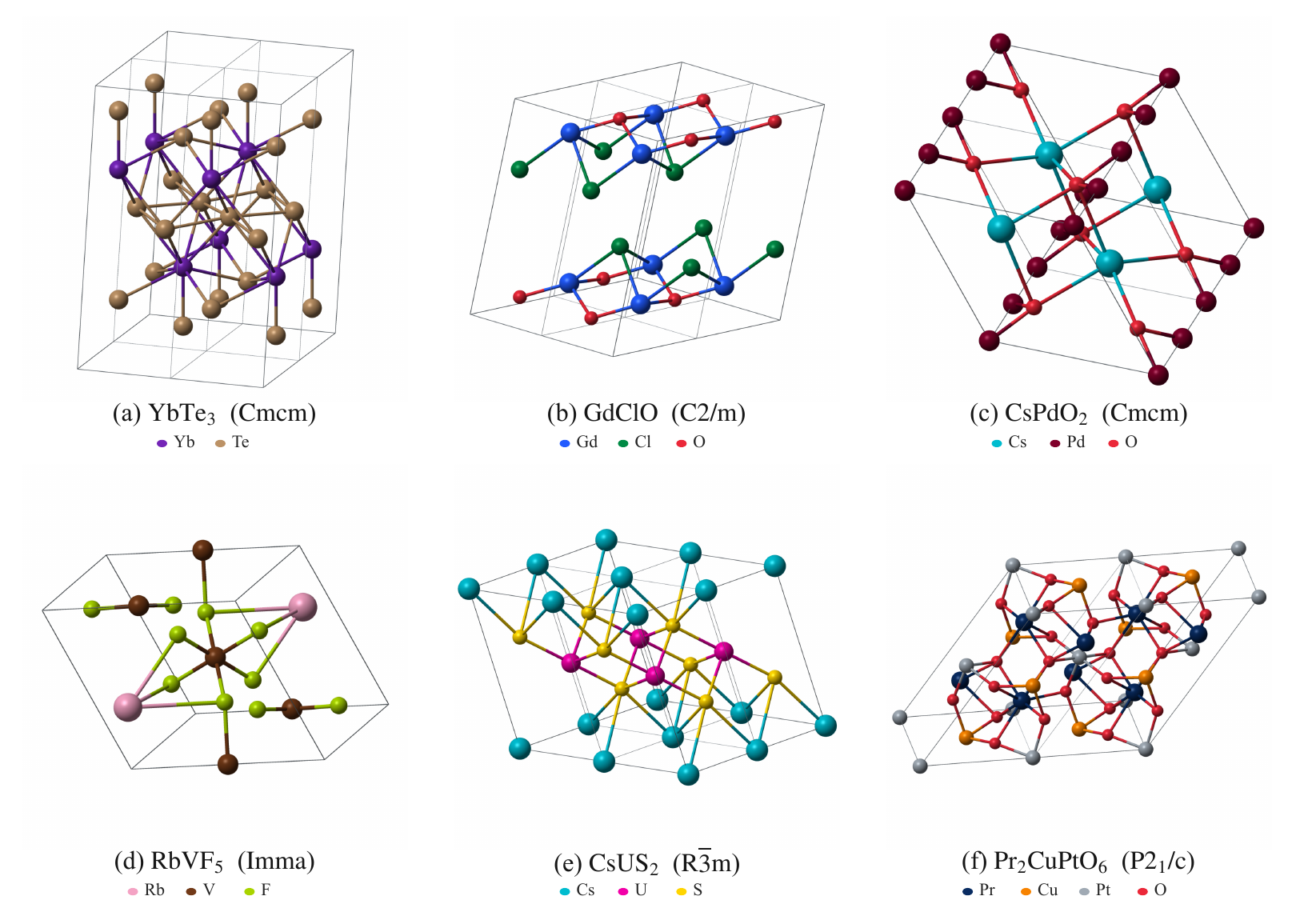}
\caption{Example crystals generated by \ours{} on MP-20, shown before relaxation, with their chemical formulae and space groups. Colors distinguish elements.}
\label{fig:generated-crystal-examples}
\end{figure}

\end{document}